\documentclass[reqno]{amsart}

\usepackage{booktabs}
\usepackage{IEEEtrantools}
\usepackage{amssymb,latexsym,amsfonts,amsmath}
\usepackage{mathrsfs}
\usepackage{graphicx}
\usepackage{dsfont}
\usepackage{tikz}
\usetikzlibrary{calc,shapes,arrows}

\usepackage{algorithm}
\usepackage{algorithmic}

\usepackage{xspace}

\newtheorem{theorem}{Theorem}[section]
\newtheorem{lemma}[theorem]{Lemma}

\newtheorem{definition}[theorem]{Definition}

\newtheorem{remark}[theorem]{Remark}
\newtheorem{assumption}{Assumption}
\numberwithin{equation}{section}

\newcommand{\R}{{\mathbb{R}}}

\newcommand{\N}{{\mathbb{N}}}

\newcommand{\Let}{:=}
\newcommand{\EE}{\mathds{E}}
\newcommand{\PP}{\mathds{P}}

\usepackage{fancyhdr}

\newenvironment{nouppercase}{%
	\renewcommand{\uppercasenonmath}[1]{}}{}

\begin{document}

\begin{abstract}
In this work, we propose a compositional data-driven approach for the formal estimation of collision risks for autonomous vehicles (AVs) with black-box dynamics while acting in a stochastic multi-agent framework. The proposed approach is based on the construction of sub-barrier certificates for each stochastic agent via a set of data collected from its trajectories while providing an a-priori guaranteed confidence on the data-driven estimation. In our proposed setting, we first cast the original collision risk problem for each agent as a robust optimization program (ROP). Solving the acquired ROP is not tractable due to an unknown model that appears in one of its constraints. To tackle this difficulty, we collect finite numbers of data from trajectories of each agent and provide a scenario optimization program (SOP) corresponding to the original ROP. We then establish a probabilistic bridge between the optimal value of SOP and that of ROP, and accordingly, we formally construct a sub-barrier certificate for each unknown agent based on the number of data and a required level of confidence. We then propose a compositional technique based on small-gain reasoning to quantify the collision risk for multi-agent AVs  with some desirable confidence based on sub-barrier certificates of individual agents constructed from data. For the case that the proposed compositionality conditions are not satisfied, we provide a relaxed version of compositional results without requiring any compositionality conditions but at the cost of providing a potentially conservative collision risk. Eventually, we also present our approaches for \emph{non-stochastic} multi-agent AVs. We demonstrate the effectiveness of our proposed results by applying them to a vehicle platooning consisting of $100$ vehicles with $1$ leader and $99$ followers. We formally estimate the collision risk for the whole network by collecting sampled data from trajectories of each agent.
\end{abstract}

\title{{\LARGE Formal Estimation of Collision Risks for Autonomous Vehicles: A Compositional Data-Driven Approach}$^*$\footnote[1]{$^*$This work was supported in part by the Swiss Reinsurance Company, Ltd.}}

\author{{\bf {\large Abolfazl Lavaei}}$^1$}
\author{{\bf {\large Luigi Di Lillo}}$^2$}
\author{{\bf {\large Andrea Censi}}$^1$}
\author{{\bf {\large Emilio Frazzoli}}$^1$\\
	{\normalfont $^1$Institute for Dynamic Systems and Control, ETH Zurich, Switzerland}\\
    {\normalfont $^2$Property and Casualty Solutions, Reinsurance, Swiss Reinsurance Company, Ltd., Zurich, Switzerland}\\
\texttt{\{alavaei,acensi,efrazzoli\}@ethz.ch}, \texttt{Luigi\_DiLillo@swissre.com}}

\pagestyle{fancy}
\lhead{}
\rhead{}
  \fancyhead[OL]{Abolfazl Lavaei, Luigi Di Lillo, Andrea Censi, Emilio Frazzoli}

  \fancyhead[EL]{Formal Estimation of Collision Risks for Autonomous Vehicles: A Compositional Data-Driven Approach} 
  \rhead{\thepage}
 \cfoot{}
 
\begin{nouppercase}
	\maketitle
\end{nouppercase}

\section{Introduction}

In the near future, we expect to see fully autonomous vehicles (AVs), aircrafts, and robots, all of which should be able to make their own decisions without direct human involvement. Although this technology theme for AVs provides many potential advantages, \emph{e.g.,} fewer traffic collisions, reduced traffic congestion, increased roadway capacity, relief of vehicle occupants from driving, etc., there have been at least two
critical challenges that need to be considered. First and foremost, closed-form mathematical models for many complex and heterogeneous physical systems, including AVs, are either not available or equally complex to be of any practical use. Accordingly, one cannot employ model-based techniques to analyze this type of complex unknown systems. Although there are some model identification techniques available in the relevant literature to learn an approximate model from data, \emph{e.g.,}~\cite[and references herein]{Hou2013model}, acquiring an accurate model for complex systems is always complicated, time-consuming and expensive. As the second difficulty,
providing safety certification and guaranteeing correctness of the design
of such AVs in a formal as well as time- and cost-effective way have been always the major obstacles in their successful deployment.
These main challenges motivated us to develop a compositional data-driven approach to bypass the system identification phase and directly evaluate the AVs performance based on data collected from trajectories of unknown agents.

\begin{figure*}
	\begin{center}
		\includegraphics[width=15cm]{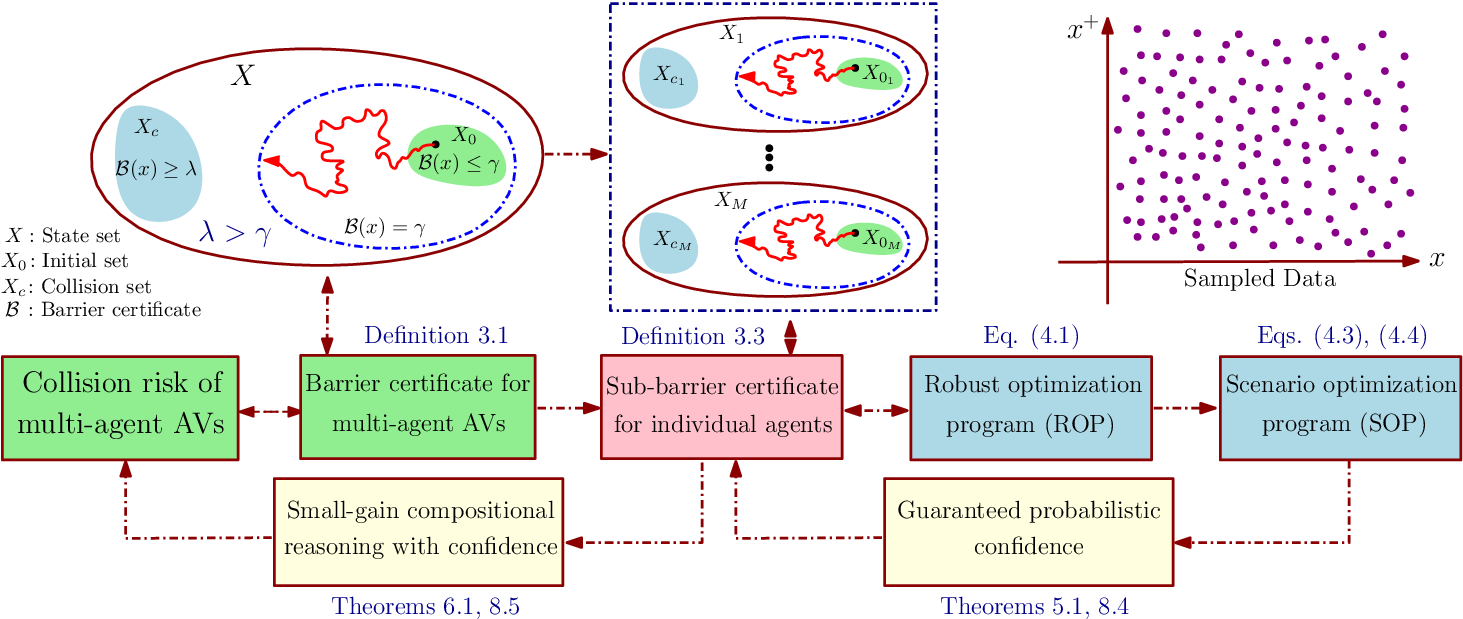}
		\caption{A graphical representation of the article's structure and contributions.}
		\label{Diagram}
	\end{center}
\end{figure*}

Over the past few years, a number of advances have been made in developing various techniques for evaluating AVs performance. Such techniques have mainly relied on methods from artificial intelligence (AI), machine learning, control theory, and optimization, \emph{e.g.,}~\cite{chouhan2020formal,paigwar2020probabilistic,barbier2019validation,an2020uncertainty,barbot2017statistical}, to name a few. Another promising approach for the safety verification of complex dynamical systems is to employ \emph{barrier certificates}, initially introduced in \cite{prajna2004safety,prajna2007framework}. This approach has received significant attentions in the past decade, as a \emph{discretization-free technique}, for formal verification and synthesis of non-stochastic \cite{borrmann2015control,wang2017safety,ames2019control,lopez2020robust,folkestad2020data,robey2021learning,verginis2021safety}, and stochastic dynamical systems \cite{zhang2010safety,yang2020efficient,M.Ahmadi,ahmadi2019safe,santoyo2019verification,clark2019control,Pushpak2019,Lavaei_TAC22,AmyAutomatica2020_J,Amy_LCSS20,Niloo_TCNS,Lavaei_ACC2022,Lavaei_Survey}, to name a few. In particular, barrier certificates are Lyapunov-like functions defined over the state space of the system to enforce a set of inequalities on both the function itself and its one-step transition. An appropriate level set of a barrier certificate separates an unsafe region, we call it here \emph{collision set}, from all system trajectories starting from a given set of initial states. Consequently, the existence of such a function provides a formal probabilistic certificate for the safety of the system (cf. Fig.~\ref{Diagram}).

Our main contribution in this work is to propose, for the first time, a compositional data-driven scheme for the formal estimation of collision risks for stochastic multi-agent AVs with black-box dynamics. Our proposed approach is based on the construction of sub-barrier certificates for each agent via a set of data collected from its trajectories. To do so, we first reformulate the original collision risk problem as a robust optimization program (ROP). Since the proposed ROP is not tractable due to an unknown model appearing in its constraints, we collect finite numbers of data from trajectories of each agent and provide a scenario optimization program (SOP) corresponding to the original ROP. We then build a probabilistic relation between the optimal value of SOP and that of ROP, and as a result, we formally construct a sub-barrier certificate for each agent based on the number of data and a required level of confidence. We propose a compositional technique based on small-gain conditions to quantify the collision risk for the multi-agent AV via sub-barrier certificates of individual agents constructed from data. We also propose a relaxed version of compositional results without requiring any compositionality condition but at the cost of providing a potentially conservative collision risk. Finally, we also present our proposed approaches for \emph{non-stochastic} multi-agent AVs. We demonstrate the effectiveness of our proposed techniques by applying them to a vehicle platooning consisting of $100$ vehicles with $1$ leader and $99$ followers (presented as a running example). Proofs of all statements are provided in Appendix. A graphical representation of the structure of the article and its contributions is illustrated in Fig.~\ref{Diagram}.

Data-driven safety verification of stochastic systems via barrier certificates is also studied in~\cite{Ali_ADHS21}. Our proposed approach here differs from the one in~\cite{Ali_ADHS21} in three main directions. First and foremost, we propose here a compositional data-driven scheme based on small-gain reasoning for the formal estimation of collision risks for multi-agent AVs, whereas the results in~\cite{Ali_ADHS21} only deal with monolithic systems, and accordingly, they are not practical in the case of facing high-dimensional systems. Moreover, we propose here a relaxed compositional approach in the case that our compositionality conditions are not fulfilled. Second, we propose our approaches for both \emph{stochastic and non-stochastic} large-scale systems, while the results of~\cite{Ali_ADHS21} are only tailored to stochastic systems. As the last main contribution, in order to propose our compositional results, we deal here with a non-convex class of robust and scenario optimization programs,  whereas the proposed results in~\cite{Ali_ADHS21} are only applicable to convex optimization problems.

The rest of the paper is organized as follows. Section~\ref{Sec: Problem} gives mathematical preliminaries and notation, and formal definitions of stochastic (multi)agent AVs. Section~\ref{SBC} provides notions of (sub)barrier certificates. Section~\ref{Sec: Data_SBC} is dedicated to data-driven construction of sub-barrier certificates. In Section~\ref{Sec: Data}, we establish a formal relation between the optimal value of SOP and that of ROP. Section~\ref{Sec:Com} contains the main compositionality results for multi-agent AVs based on small-gain reasoning. In Section~\ref{Sec: Com_Relaxed}, we propose a relaxed version of compositional results in which the compositionality conditions are no longer required. We present our results for \emph{non-stochastic} multi-agent AVs in Section~\ref{Sec: Deterministic}.

\section{Problem Description}\label{Sec: Problem}

\subsection{Preliminaries and Notation}
We consider a probability space $(\Omega,\mathcal F_{\Omega},\mathds{P}_{\Omega})$, where $\Omega$ is the sample space,
$\mathcal F_{\Omega}$ is a sigma-algebra on $\Omega$ comprising subsets of $\Omega$ as events, and $\mathds{P}_{\Omega}$ is a probability measure assigning probabilities to events. We assume that random variables introduced in this article are measurable functions of the form $X:(\Omega,\mathcal F_{\Omega})\rightarrow (\mathcal V_X,\mathcal F_X)$. Given the probability space $(\Omega,\mathcal F_{\Omega},\mathds{P}_{\Omega})$, we denote the $N$-Cartesian product set of $\Omega$ by $\Omega^N$, and its corresponding product measure by $\PP^N$. A random variable $\varsigma$ with standard normal distributions (zero mean and variance of identity) is denoted by $\varsigma(\cdot) \sim\mathcal N(0, \mathds{I}_n)$. A topological space $S$ is called a Borel space if it is homeomorphic to a Borel subset of a Polish space (\emph{i.e.,} a separable and completely metrizable space). Any Borel space $S$ is assumed to be endowed with a Borel sigma-algebra, which is denoted by $\mathds B(S)$.

We denote the set of real, positive and non-negative real numbers by $\mathbb{R},\mathbb{R}^+$, and $\mathbb{R}^+_0$, respectively. $\mathbb{N} := \{0,1,2,...\}$ represents the set of non-negative integers and $\mathbb{N}_{\geq 1}=\{1,2,...\}$ is the set of positive integers. Given $M$ vectors $x_i \in \mathbb{R}^{n_i}$, $x=[x_1;...;x_M]$ denotes the corresponding column vector of dimension $\sum_i n_i$. Given a symmetric matrix $A$, $\lambda_{\max}(A)$ denotes the maximum eigenvalue of $A$. Given any $a\in\mathbb R$, $\vert a\vert$ denotes the absolute value of $a$. Given a vector $x\in\mathbb{R}^{n}$, $\Vert x\Vert$ denotes the Euclidean norm of $x$. For any matrix $P\in\mathbb R^{m\times n}$, we have $\|P\| := \sqrt{\lambda_{\max}(P^\top P)}$. We denote the indicator function of a subset $\mathcal A$ of a set $X$ by $\mathbf 1_{\mathcal A}:X\rightarrow\{0,1\}$, where $\mathbf 1_{\mathcal A}(x)=1$ if and only if $x\in {\mathcal A}$, and $0$ otherwise. Symbol $\mathds{1}_n$ denotes a column vector in $\mathbb R^{n\times{1}}$ with all elements equal to one.  If a system $\mathcal A$ satisfies a property $\varphi$, we denote it by $\mathcal A \models \varphi$. We also use $\models$ to show the \emph{feasibility} of a solution for an optimization problem. 
Given functions $f_i:X_i\rightarrow Y_i$, for any $i\in\{1,\ldots,M\}$, their Cartesian product $\prod_{i=1}^{M}f_i:\prod_{i=1}^{M}X_i\rightarrow\prod_{i=1}^{M}Y_i$ is defined by $(\prod_{i=1}^{M}f_i)(x_1,\ldots,x_M)=[f_1(x_1);\ldots;f_M(x_M)]$.  Given a measurable function $f:\mathbb N\rightarrow\mathbb{R}^n$, the (essential) supremum of $f$ is denoted by $\Vert f\Vert_{\infty} \Let \text{(ess)sup}\{\Vert f(k)\Vert,k\geq 0\}$. A function $\phi: \mathbb{R}_{\geq 0} \rightarrow \mathbb{R}_{\geq 0}$ is said to be a class $\mathcal{K}$ function if it is continuous, strictly increasing, and $\phi(0)=0$. A class $\mathcal{K}$ function $\phi(\cdot)$ belongs to class $\mathcal{K}_\infty$ if $\phi(s) \rightarrow \infty$ as $s \rightarrow \infty$.

\subsection{Stochastic (Multi-)Agent AVs}
In this work, we consider dynamics of each agent as a discrete-time \emph{stochastic} system, as formalized in the following definition.

\begin{definition}\label{Def:1}
	Each agent of AVs is a tuple 
	\begin{equation}\label{Eq:1}
	\mathcal A_i=(X_i,W_i,\varsigma_i,f_i),
	\end{equation}
	where:
	\begin{itemize}
		\item
		$X_i\subseteq \mathbb R^{n_i}$ is a Borel space as the state set of the
		agent;
		\item $W_i\subseteq \mathbb R^{p_i}$ is a Borel space as the interaction set of the agent;
		\item
		$\varsigma_i$ is a sequence of independent-and-identically distributed
		(i.i.d.) random variables from the sample space $\Omega_i$ to a measurable space $(\mathcal{V}_{\varsigma_i},\mathcal F_{\varsigma_i})$, \emph{i.e.,}
		$$
		\varsigma_i:=\{\varsigma_i(k):(\Omega_i,\mathcal F_{\Omega_i})\rightarrow (\mathcal{V}_{\varsigma_i},\mathcal F_{\varsigma_i}),\,\,k\in\N\}; 
		$$
		\item$f_i:X_i\times W_i \times \mathcal V_{\varsigma_i} \rightarrow X_i$ is a measurable function characterizing the state evolution of $\mathcal A_i$.
	\end{itemize}
\end{definition} 
The evolution of the state of $\mathcal A_i$ for a given initial state $x_{0_i} = x_i(0)\in
X_i$ and an interaction sequence $\{w_i(k):\Omega_i\rightarrow W_i,\,\,k\in\mathbb N\}$ is
described as
\begin{equation}\label{Eq:2}
\mathcal A_i\!:x_i(k+1)=f_i(x_i(k),w_i(k),\varsigma_i(k)),
\quad k\in\mathbb N.
\end{equation}
We call the random sequence $x_{x_{0_i}w_i}\!\!:\Omega_i \times\mathbb N \rightarrow X_i$ satisfying~\eqref{Eq:2} the \textit{solution process} of $\mathcal A_i$ under an interaction $w_i$ and an initial state $x_{0_i}$.

Since the main goal in this work is to formally estimate the collision risks for AVs, we assume that the controller for each agent is already designed and deployed to the system. There is no
requirement on the type of controllers in our setting. We model the effect of other agents, influencing the state of the current agent, via interactions $w_i$ which can be also captured by the deployed closed-loop controller inside AVs. In the following, we provide a formal definition of multi-agent AVs without interactions $w_i$ that is considered as a composition of several agents with interactions. 

\begin{definition}\label{Def:2}
	Consider $M\in\mathbb N_{\geq1}$ agents $\mathcal A_i=(X_i,W_i,\varsigma_i,f_i)$, $i\in \{1,\dots,M\}$, with their interactions partitioned as	
	\begin{align}\label{Eq:3}
	w_i&=[{w_{i1};\ldots;w_{i(i-1)};w_{i(i+1)};\ldots;w_{iM}}], \quad w_{ij}\in W_{ij},
	\end{align}
	where $W_i:=\prod_{j}W_{ij}$. A multi-agent AV is defined as
	$\mathcal A=(X,\varsigma,f)$, denoted by
	$\mathcal{I}(\mathcal A_1,\ldots,\mathcal A_M)$, such that $X:=\prod_{i=1}^{M}X_i$,  $\varsigma = [\varsigma_1;\dots;\varsigma_M]$, and $f:=\prod_{i=1}^{M}f_{i}$ subjected to the following interconnection constraint:
	\begin{align}\label{Eq:4}
	\forall i,j\in \{1,\dots,M\},i\neq j\!: ~~ w_{ij}=x_{j}, ~~ X_{j}\subseteq W_{ij}.
	\end{align}
	Such a multi-agent AV can be represented by
	\begin{equation}\label{Eq:5}
	\mathcal A\!:x(k+1)=f(x(k),\varsigma(k)), ~ k\in\mathbb N, \quad \text{with}~ f: X \times \mathcal V_{\varsigma} \rightarrow X.
	\end{equation}
\end{definition}

For the sake of a better illustration of the results, we provide our case study as a running example throughout the paper.

\begin{figure}\vspace{-0.3cm}
	\centering
	\includegraphics[scale=0.13]{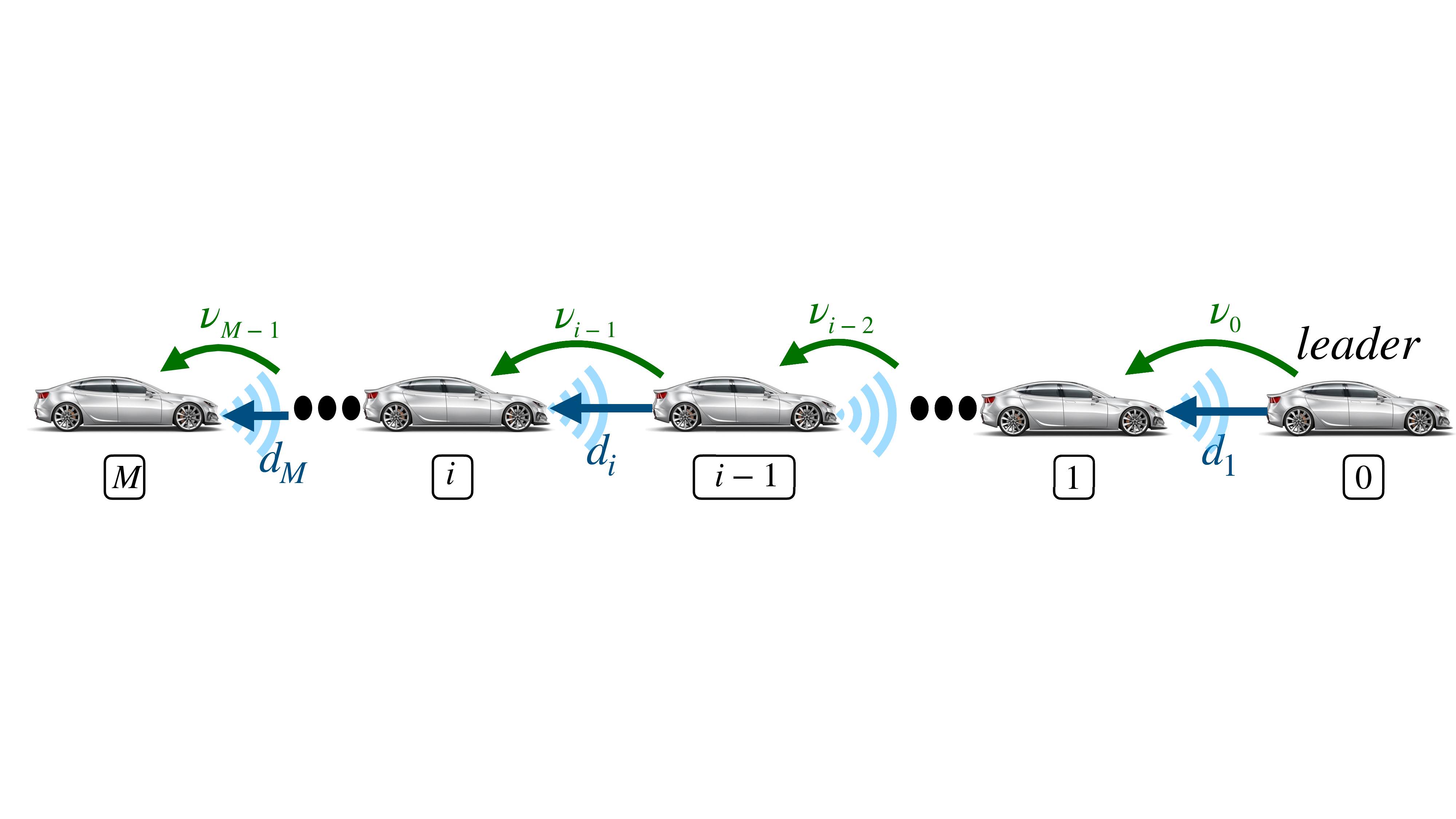} \vspace{-2.2cm}
	\caption{Vehicle platooning in a network of $100$ vehicles.}
	\label{fig:cars}
\end{figure}

{\bf Case Study: Vehicle Platooning.} Consider a vehicle platooning in a network of $100$ vehicles with $1$ leader and $99$ followers $($see Fig.~\ref{fig:cars}$)$. The model of this case study is adapted from~\cite{sadraddini2017provably}. The evolution of states can be described by the following multi-agent AV:
\begin{equation*}
\mathcal A\!: x(k+1)=Ax(k)+u(k)+R\varsigma(k),
\end{equation*}
where $A$ is a block matrix with diagonal blocks $A_x$, and off-diagonal blocks $A_{i(i-1)}=A_w, i\in\{2,\ldots,M\}$, where
\[A_x=\begin{bmatrix}
1 & -1\\ 0  &1
\end{bmatrix}\!\!, \quad A_w=\begin{bmatrix}
0 & \varpi\\ 0& 0
\end{bmatrix}\!\!,
\]
with $\varpi=0.01$ being the interconnection degree, and all other off-diagonal blocks are zero matrices of appropriate dimensions. Moreover, $R$ is a partitioned matrix with main diagonal blocks $\bar R=\begin{bmatrix}
0.03 & 0\\ 0 & 0.06
\end{bmatrix}\!\!,$ and all other off-diagonal blocks being zero matrices of appropriate dimensions. Furthermore, $x(k)=[x_1(k);\ldots;x_M(k)]$, $\varsigma(k)=[\varsigma_{1}(k);\ldots;\varsigma_M(k)]$, and $u(k)=[u_1(k);\ldots;u_M(k)]$, where $u_i(k)$ is the control input for each agent. By considering each individual agent $\mathcal A_i$ described as
\begin{equation*}  
\mathcal A_i\!: x_i(k+1)=A_xx_i(k)+ u_i(k)+A_ww_i(k)+\bar R\varsigma_{i}(k),
\end{equation*}
one can readily verify that $\mathcal A=\mathcal{I}(\mathcal A_1,\ldots,\mathcal A_M)$, where $w_i(k)=[0;w_{i(i-1)}(k)],i\in\{1,\ldots,M\}$, (with $w_{i(i-1)}=x_{i-1},w_{1,0}=0$). The state of the $i$-th vehicle is defined as $x_i=[d_i;\nu_i]$, for any $i\in\{1,\ldots,M\}$, where $d_i$ denotes the relative distance between the vehicle $i$ and its proceeding vehicle $i-1$ (the $0$-$th$ vehicle represents the leader), and $\nu_i$ is its velocity in the leader's frame. The overall control objective in vehicle platooning is for each vehicle to adjust its speed in order to maintain a safe distance from the vehicle ahead. We assume that the controller for each agent is already designed and deployed as $u_i= [-0.2d_i + 1.1\nu_i + 0.1 ; 0.01d_i - 0.9\nu_i]$, for any $i\in\{1,\ldots,M\}$. 

In the next section, in order to estimate the collision risk for multi-agent AVs
in finite time horizons, we present notions of barrier and sub-barrier certificates for, respectively, multi-agent AVs $\mathcal A$ (without interactions) and individual agents $\mathcal A_i$ (with interactions). 

\section{(Sub-)Barrier Certificates}\label{SBC}
In this section, we first present the notion of barrier certificates for  multi-agent AVs $\mathcal A$ without interactions, as the following definition. 
\begin{definition}\label{Def:3}
	Given the multi-agent AV $\mathcal A = (X,\varsigma,f)$, a non-negative function $\mathcal{B}\!:X\rightarrow\mathbb{R}_{0}^{+}$ is called a barrier certificate (BC) for $\mathcal A$ if there exist $\gamma, \lambda \in\mathbb R^+$, with $\lambda > \gamma$, $0<\kappa <1$, and $\psi\in \mathbb R_{0}^+$, such that:
	\begin{align}\label{Eq:6_1}
	&\mathcal{B}(x)\leq\gamma, \quad\quad\quad\quad\forall x\in X_{0},\\\label{Eq:6_2}
	&\mathcal{B}(x)\geq\lambda, \quad\quad\quad\quad\forall x\in X_c,\\\label{Eq:6_3}
	&\EE\Big[\mathcal{B}(f(x,\varsigma))\,\big |\, x\Big]\leq\kappa\mathcal{B}(x)+\psi,\quad\quad\quad\forall x \in X,
	\end{align}
	where $X_{0},X_c\subseteq X$ are initial and collision sets, respectively. 
\end{definition} 

Given a collision event $\varphi= (X_{0},X_c,\mathcal T)$, the multi-agent AV $\mathcal A$ may have a collision, denoted by $\mathcal A\models_{\mathcal T} \varphi,$ if a trajectory of $\mathcal A$ starting from the initial set $X_{0}$ reaches the collision set $X_c$ within the time horizon $\mathcal{T}$. 
Since trajectories of the multi-agent AV are probabilistic, we are interested in computing the collision risk $\PP\{\mathcal A\models_{\mathcal T} \varphi\}$.
Now we employ Definition~\ref{Def:3} and quantify the collision risk for the multi-agent AVs in~\eqref{Eq:5} via the next theorem~\cite{1967stochastic}.

\begin{theorem}\label{Thm:1}
	Consider the multi-agent AV $\mathcal A$ defined in Definition~\ref{Def:2} and a finite time horizon $\mathcal T\in \mathbb{N}$. Suppose there exists a non-negative barrier certificate $\mathcal B$ satisfying conditions~\eqref{Eq:6_1}-\eqref{Eq:6_3}. Then the collision risk for the multi-agent AV $\mathcal A$ is bounded by
	\begin{equation}\label{Eq:7}
	\PP\Big\{\mathcal A\models_{\mathcal T} \varphi\Big\}\leq\begin{cases} 
	1-(1-\frac{\gamma}{\lambda})(1-\frac{\psi}{\lambda})^{\mathcal T}\!\!, & \quad\quad\text{if } \lambda \geq \frac{\psi}{1-\kappa}, \\
	(\frac{\gamma}{\lambda})\kappa^{\mathcal T}+(\frac{\psi}{(1-\kappa)\lambda})(1-\kappa^{\mathcal T}), & \quad\quad\text{if } \lambda< \frac{\psi}{1-\kappa}.  \\
	\end{cases}
	\end{equation}
\end{theorem}

In general, searching for barrier certificates for multi-agent AVs as in Definition~\ref{Def:3} is computationally very expensive, even if the model is known, mainly due to the high-dimension of the underlying system. Accordingly, we present in the following a definition of sub-barrier certificates (SBC) for individual agents $\mathcal A_i$ and propose in Section~\ref{Sec:Com} a compositional approach based on small-gain reasoning to construct a BC of multi-agent AVs based on SBC of individual agents.

\begin{definition}\label{Def:4}
	Given an agent $\mathcal A_i = (X_i,W_i,\varsigma_i,f_i)$, a non-negative function $\mathcal{B}_i\!:X_i\rightarrow\mathbb{R}_{0}^{+}$ is called a sub-barrier certificate (SBC) for $\mathcal A_i$ if there exist $\gamma_i, \lambda_i\in\mathbb R^+$, $\alpha_i,\rho_i,\psi_i\in\mathbb R_0^+$, and $0<\kappa_i <1$, such that:
	\begin{align}\label{Eq:8_1}
	&\mathcal B_i(x_i) \geq \alpha_i\Vert x_i\Vert^2,\quad\quad\quad\quad\forall x_i \in X_i,\\\label{Eq:8_2}
	&\mathcal{B}_i(x_i)\leq \gamma_i,\quad\quad\quad\quad\quad\quad\quad\!\!\!\forall x_i\in X_{0_i},\\\label{Eq:8_3}
	&\mathcal{B}_i(x_i)\geq \lambda_i, \quad\quad\quad\quad\quad\quad\quad\!\!\!\forall x_i\in X_{c_i},\\\label{Eq:8_4}
	& \EE\Big[\mathcal{B}_i(f_i(x_i,w_i,\varsigma_i))\,\big |\, x_i,w_i\Big]\leq\kappa_i\mathcal{B}_i(x_i)+\rho_i \Vert w_i\Vert^2 + \psi_i, \quad\quad\quad\!\!\!\forall x_i \in X_i, ~\forall w_i \in W_i,
	\end{align}
	where $X_{0_i}, X_{c_i}\subseteq X_i$ are initial and collision sets of agents, respectively. 
\end{definition}

\begin{remark}
	Note that $\alpha_i$ and $\rho_i$ in Definition~\ref{Def:4} could be, without loss of generality, $\mathcal{K}_\infty$ functions as proposed in~\cite{Lavaei_IFAC20}. We allow them in this work to be linear for the sake of an easier presentation. 
\end{remark}

{\bf Case Study: Vehicle Platooning (continued).} Regions of interest for each vehicle are given as $X_i = [0,0.7]\times[-3.55,0.2], X_{0_i} = [0.35,0.7]\times[-0.9,0.2]$, and $X_{c_i} = [0,0.3]\times[-3.55,-2.9]$, for all $i\in\{1,\ldots,M\}$. We fix the structure of our sub-barrier certificates as $\mathcal B_i(q_i,x_i) = q_{i_1}d_i^2 + q_{i_2} d_i^4 + q_{i_3} d_i\nu_i + q_{i_4}\nu_i^4 + q_{i_5}\nu_i^2$, for all  $i\in\{1,\ldots,100\}$, with unknown coefficients $q_{i_1}$\!-$q_{i_5}$.

In the next section, we provide our data-driven scheme for the construction of SBC for each agent. 

\section{Data-Driven Construction of SBC}\label{Sec: Data_SBC}

In this section, we assume that the transition map $f_i$ and the distribution of $\varsigma_i$ in~\eqref{Eq:2} are both unknown, and we employ the term \emph{black-box} models to refer to this type of systems. We fix the structure of SBC as $\mathcal{B}_i(q_i,x_i)=\sum_{j=1}^{r} q_{i_j}\bar p_{i_j}(x_i)$ with some user-defined (possibly nonlinear) basis functions $\bar p_{i_j}(x_i)$ and unknown coefficients $q_i=[q_{i_1};\ldots;q_{i_r}] \in \mathbb{R}^r$. It is worth mentioning that our proposed
techniques do not put any restrictions on the type of basis
functions in the structure of SBC. For instance, in the case of having polynomial-type barrier certificates, basis functions $\bar p_{i_j}(x_i)$ are monomials over $x_i$.

In order to enforce conditions~\eqref{Eq:8_1}-\eqref{Eq:8_4} in Definition~\ref{Def:4}, we first cast our collision risk problem for each agent as the following robust optimization program (ROP), for any $i\in \{1,\dots,M\}$:

\begin{align}\label{ROP}
\textbf{ROP}\!:\left\{\hspace{-0.15cm}
\begin{array}{ll}
\underset{[\Theta_i;\eta_i]}{\textbf{min}}\quad \eta_i\quad \quad\\
~\!\textbf{s.t.}\quad\,\,\,\,\,\!
\max\Big\{g_{i_z}(x_i,\Theta_i),g_{i_5}(x_i,w_i,\Theta_i)\Big\} \leq \eta_i, \\
\qquad\,\,\,\,\,\,\,\,\,z\in\{1,\dots,4\},\forall x_i \in X_i, \forall w_i \in W_i,
\end{array}
\right.
\end{align}
where:
\begin{align}\notag
&\Theta_i=[\gamma_i;\lambda_i;\alpha_i;\kappa_i;\rho_i;\psi_i;{q_{i_1};\dots;q_{i_r}}],\\\notag
&\eta_i\in\R, \gamma_i,\lambda_i\in\mathbb{R}^+,\alpha_i,\rho_i,\psi_i\in\mathbb R_0^+, \kappa_i\in(0,1),\\\notag
&g_{i_1}\!=-\mathcal{B}_i(q_i,x_i),~ g_{i_2} = \alpha_i\Vert x_i\Vert^2 -\mathcal{B}_i(q_i,x_i),\\\notag &g_{i_3}\!=(\mathcal{B}_i(q_i,x_i)-\gamma_i)\mathbf{1}_{X_{0_i}}(x_i),g_{i_4}\!=(\lambda_i-\mathcal{B}_i(q_i,x_i))\mathbf{1}_{X_{c_i}}\!(x_i), \\\label{Eq:10}
&g_{i_5}\!=\EE\Big[\mathcal{B}_i(q_i,f_i(x_i,w_i,\varsigma_i))\,\big |\, x_i, w_i\Big]-\kappa_i\mathcal{B}_i(q_i,x_i)-\rho_i\Vert w_i\Vert^2 -\psi_i,
\end{align}
with $\mathbf{1}_{X_{0_i}}(x_i)$ and $\mathbf{1}_{X_{c_i}}(x_i)$ being indicator functions acting on initial and collision sets, respectively. We denote the optimal value of ROP by $\eta_{R_i}^*$. If $\eta_{R_i}^* \leq 0$, a solution to the ROP implies the satisfaction of conditions~\eqref{Eq:8_1}-\eqref{Eq:8_4} in Definition~\ref{Def:4}.

To solve the proposed ROP in~\eqref{ROP}, we face two major difficulties. First, the proposed ROP in~\eqref{ROP} has infinitely-many constraints since the state and interaction of $\mathcal A_i$ both live in continuous sets (\emph{i.e.,} $x_i \in X_i$, $w_i \in W_i$). In addition and more importantly, one needs to know the precise map $f_i$ in $g_{i_5}$ in order to solve the optimization problem in~\eqref{ROP}. To tackle these two difficulties, we propose a scenario optimization program corresponding to the original ROP. Suppose $(\hat x_{i_l},\hat w_{i_l})^{N_i}_{l=1}$ denote $N_i$ i.i.d. sampled data within $X_i \times W_i$. We take two consecutive sampled data-points from trajectories of $\mathcal A_i$ as the pair of $(x_i(k),x_i(k+1))$ and denote it by $(\hat x_i, f_i(\hat x_i, \hat w_i,\varsigma_i))$. We now propose the following scenario optimization program (SOP), for any $i\in \{1,\dots,M\}$: 

\begin{align}\label{SOP_1}
\textbf{SOP}_\text{N}\!:\left\{\hspace{-0.15cm}
\begin{array}{ll}
\underset{[\Theta_i;\eta_i]}{\textbf{min}}\quad \eta_i\quad \quad\\
~\!\textbf{s.t.}\quad\,\,\,~
\max\Big\{g_{i_z}(\hat{x}_{i_l},\Theta_i),g_{i_5}(\hat{x}_{i_l},\hat{w}_{i_l},\Theta_i)\Big\} \leq\eta_i, \\
\qquad \,\,\,\,\,\,\,\,\,~ z \in\{1,\dots,4\},\forall \hat{x}_{i_l} \in X_i, \forall \hat{w}_{i_l} \in W_i,\\
\qquad \,\,\,\,\,\,\,\,\,~\forall l\in \{1,\cdots,N_i\},
\end{array}
\right.
\end{align}
where: 
\begin{align*}
&\Theta_i=[\gamma_i;\lambda_i;\alpha_i;\kappa_i;\rho_i;\psi_i;{q_{i_1};\dots;q_{i_r}}],\\\notag
&\eta_i\in\R, \gamma_i,\lambda_i\in\mathbb{R}^+,\alpha_i,\rho_i,\psi_i\in\mathbb R_0^+, \kappa_i\in(0,1),
\end{align*}
and $g_{i_1}$-$g_{i_5}$ are the functions as defined in~\eqref{Eq:10}. We denote the optimal value of $\text{SOP}_\text{N}$ by $\eta_{N_i}^*$. Although the problems of infinitely-many constraints and unknown map $f_i$ are resolved in $\text{SOP}_\text{N}$, there is still no closed-form solution for the expected value in $g_{i_5}$. Hence, we employ an empirical approximation of the expected value  and propose a new scenario optimization problem, denoted by SOP$_\varsigma$, for any $i\in \{1,\dots,M\}$:

\begin{align}\label{SOP_2}
\textbf{SOP}_\varsigma\!:\left\{\hspace{-0.15cm}
\begin{array}{ll}
\underset{[\Theta_i;\eta_i]}{\textbf{min}}\quad \eta_i\quad \quad\\
~\!\textbf{s.t.}\quad\,\,~
\max\Big\{g_{i_z}(\hat{x}_{i_l},\Theta_i),\bar g_{i_5}(\hat{x}_{i_l},\hat{w}_{i_l},\Theta_i)\Big\} \leq \eta_i,\\
\qquad \,\,\,\,\,\,\,\,~ z \in\{1,\dots,4\},\forall \hat{x}_{i_l} \in X_i, \forall \hat{w}_{i_l} \in W_i,\\
\qquad \,\,\,\,\,\,\,\,~\forall l\in \{1,\cdots,N_i\},
\end{array}
\right.
\end{align}
with
\begin{align}\notag
&\hspace{-2.7cm}\Theta_i=[\gamma_i;\lambda_i;\alpha_i;\kappa_i;\rho_i;\psi_i;{q_{i_1};\dots;q_{i_r}}],\\\notag
&\hspace{-2.6cm}\eta_i\in\R, \gamma_i,\lambda_i\in\mathbb{R}^+,\alpha_i,\rho_i,\psi_i\in\mathbb R_0^+, \kappa_i\in(0,1),\\\label{Eq:9}
\bar{g}_{i_5}(\hat{x}_{i_l},\hat{w}_{i_l},\Theta_i)\!=&\frac{1}{\hat{N}_i}\sum_{j=1}^{\hat{N}_i}\mathcal{B}_i(q_i,f_i(\hat{x}_{i_l},\hat{w}_{i_l}, \hat\varsigma_{i_j}))-\kappa_i\mathcal{B}_i(q_i,\hat{x}_{i_l})-\rho_i \Vert \hat{w}_{i_l}\Vert^2 - \psi_i + \mu_i,
\end{align}
where $\hat{N}_i\in\mathbb{N}$ and $\mu_i\in \mathbb{R}_0^+$ are, respectively, the number of samples required for the empirical approximation and the corresponding error introduced by this approximation. We denote the optimal value of the objective function  SOP$_\varsigma$ by $\eta^*_{\varsigma_i}$. 

\begin{remark}
	Note that condition $\bar g_{i_5}$ is not convex due to a bilinearity between decision variables $q_i$ and the unknown variable $\kappa_i$. Although a mild convexity can be observed since both $\kappa_i$ and $q_i$ are
	scalar, the SOP$_\varsigma$ in~\eqref{SOP_2} is in principle non-convex. To tackle this issue, we consider $\kappa_i$ in a finite set with a cardinality $m_i$ while solving SOP$_\varsigma$ in~\eqref{SOP_2}. Accordingly, we propose our data-driven approach in Section~\ref{Sec: Data} by employing the cardinality $m_i$ in computing the minimum number of data required for solving SOP$_\varsigma$~\eqref{SOP_2} (cf. Theorem~\ref{Thm:2}). 
\end{remark}

\begin{remark}
	Since the collected data required for solving \text{SOP}$_\varsigma$~\eqref{SOP_2} should be i.i.d., one is allowed to take only one paired sample $(\hat x_i, f_i(\hat x_i, \hat w_i,\varsigma_i))$ from each trajectory of unknown systems. However, the data can be collected with any arbitrary distribution.
\end{remark}

The next lemma, borrowed from~\cite{Ali_ADHS21}, employs Chebyshev's inequality~\cite{saw1984chebyshev} to establish a relation between solutions of SOP$_\varsigma$  and SOP$_N$  with a required number of data $\hat{N}_i$, an approximation error $\mu_i$, and a desired confidence $\beta_{1_i} \in (0,1]$. 

\begin{lemma}\label{Lemma:1}
	Suppose $\hat{\mathcal{B}}_i(q_i,x_i)$ is a solution for SOP$_\varsigma$ in~\eqref{SOP_2}. For an a-priori approximation error $\mu_i\in \mathbb{R}^+$, a desired confidence $\beta_{1_i} \in (0,1]$, and an upper bound $\mathcal{Q}_i$ on the variance of the sub-barrier certificate applied on $f_i$, \emph{i.e.,} $\text{Var}\big[\mathcal{B}_i(q_i,f_i(x_i,w_i,\varsigma_i))\big]\leq \mathcal{Q}_i\in \mathbb{R}^+, \forall x_i \in X_i, w_i \in W_i$, one has
	\begin{align*}
	\PP\Big\{\hat{\mathcal{B}}_i(q_i,x_i) \models\text{SOP}_N\Big\}\geq 1- \beta_{1_i}, \quad\text{with}~ \hat{N}_i \geq \frac{\mathcal{Q}_i}{\beta_{1_i}\mu_i^2},
	\end{align*}
	for any $i\in \{1,\dots,M\}$.
\end{lemma}

\begin{remark}
	As it can be observed from Lemma~\ref{Lemma:1}, there is a relation between approximation error $\mu_i$, confidence $\beta_{1_i}$ and the required number of data $\hat{N}_i$. One can arbitrary select $\mu_i$, $\beta_{1_i}$ and compute the required number of data $\hat{N}_i$ based on Lemma~\ref{Lemma:1}. The obtained $\hat{N}_i$ is then utilized for computing the empirical approximation in~\eqref{Eq:9}.
\end{remark}
\section{Construction of SBC for Unknown Agents}\label{Sec: Data}

In this section, we aim at establishing a formal relation between the optimal value of SOP$_\varsigma$ in~\eqref{SOP_2} and that of ROP in~\eqref{ROP}, inspired by the fundamental results of~\cite{esfahani2014performance}. Accordingly, we formally quantify the SBC of unknown agents based on the number of data and a required level of confidence. To do so, we first raise the following assumption.

\begin{assumption}
	\label{Asum:1}
	Functions $g_{i_1}$-$g_{i_4}$ are all Lipschitz continuous with respect to $x_i$, and $g_{i_5}$ is Lipschitz continuous with respect to $x_i,w_i$, with, respectively, Lipschitz constants $\mathscr{L}_{g_{i_1}}\!$-$\mathscr{L}_{g_{i_4}}$, $\mathscr{L}_{g_{i_{5_k}}}\!,$ for given $\kappa_{i_k}$, where $\ k\in\{1,\dots,m_i\}$.
\end{assumption} 
We provide some explicit approaches in Lemmas~\ref{Lemma:2}, \ref{Lemma:3} to compute the required Lipschitz constants in Assumption~\ref{Asum:1}. We now propose the main result of this section.
\begin{theorem}\label{Thm:2}
	Consider an unknown agent $\mathcal A_i$ as in~\eqref{Eq:2}, and initial and collision regions $X_{0_i}$ and $X_{c_i}$, respectively. Let Assumption~\ref{Asum:1} hold. Consider the corresponding $\text{SOP}_\varsigma$ in~\eqref{SOP_2} with its associated optimal value $\eta^*_{\varsigma_i}$ and the solution $\Theta_i^* = [\gamma^*_i;\lambda^*_i;\alpha^*_i;\rho^*_i;\psi^*_i;{q}^*_{i_1};\dots;q^*_{i_r}]$, with $N_i\big(\bar\varepsilon_{2_i},\beta_{2_i}\big)$, $\bar\varepsilon_{2_i} := (\varepsilon_{2_{i_1}}, \dots, \varepsilon_{2_{i_m}})$, where 	
	\begin{align}\label{Min_Data}
	N_i(\bar\varepsilon_{2_i},\beta_{2_i}):=\min\Big\{N_i\in\N \,\big|\,\sum_{k=1}^{m_i}\sum_{j=0}^{c_i-1}\binom{N_i}{j}\varepsilon_{2_{i_k}}^j(1-\varepsilon_{2_{i_k}})^{N_i-j}\leq\beta_{2_i}\Big\},
	\end{align}
	$\beta_{2_i} \in [0,1]$, $\varepsilon_{2_{i_k}}=(\frac{\varepsilon_{1_i}}{\mathscr{L}_{g_{i_k}}})^{n_i+p_i}$, $\varepsilon_{1_i}\in [0,1] \leq \mathscr{L}_{g_{i_k}} = \max \big\{\mathscr{L}_{g_{i_1}},\mathscr{L}_{g_{i_2}},\mathscr{L}_{g_{i_3}},\mathscr{L}_{g_{i_4}},\mathscr{L}_{g_{i_{5_k}}}\big\}, k\in\{1,\dots,m_i\}$, with $n_i,p_i,c_i,m_i$ being, respectively, dimensions of state and interaction sets, the number of decision variables in $\text{SOP}_\varsigma$, and the cardinality of a finite set that $\kappa_i\in(0,1)$ takes value from it. Then if $\eta^*_{\varsigma_i} + \varepsilon_{1_i} \leq 0$, the solution $\Theta^*_i$ is a feasible solution for ROP~\eqref{ROP}, \emph{i.e.,} $\Theta^* \vDash \text{ROP}$, with a confidence of at least $1-\beta_{1_i}-\beta_{2_i}$. 
\end{theorem}

\begin{remark}
	Note that although the minimum number of data in~\eqref{Min_Data} required for solving \text{SOP}$_\varsigma$ in~\eqref{SOP_2} is exponential with respect to the dimension of unknown agents, the proposed compositional approach here significantly mitigates the computational complexity given that \text{SOP}$_\varsigma$ should be solved for each agent instead of multi-agent AVs. It may be worth mentioning that the main benefit of our techniques compared to system identification is that the proposed data-driven approach here is capable of constructing SBC for \emph{any type of nonlinear systems which are Lipschitz continuous}, whereas system identification approaches are mainly tailored to linear or some particular classes of nonlinear systems. In addition, even if one is able to find a model using system identification techniques, one still needs to search for a barrier certificate. In this case, one suffers from the computational complexity in both identifying the model as well as searching for SBC based on it.
\end{remark}
We summarize the results of Theorem~\ref{Thm:2} in Algorithm~\ref{Alg:1} to
describe the required procedure.

\begin{algorithm}[h]
	\caption{Data-driven construction of SBC for unknown agents $\mathcal A_i$}\label{Alg:1}
	\begin{center}
		\begin{algorithmic}[1]
			\REQUIRE 
			$\beta_{1_i}\in (0,1],\beta_{2_i}\in [0,1]$, $\mu_i,\mathcal{Q}_i,\mathscr{L}_{g_{i_k}} \in \mathbb{R}^+$, and degree of SBC
			\STATE
			Choose $\varepsilon_{1_i} \in [0,1]$ such that $\varepsilon_{1_i} \leq \mathscr{L}_{g_{i_k}}$	
			\STATE
			Compute $\hat{N}_i$ required for the empirical approximation in~\eqref{Eq:9} according to Lemma~\ref{Lemma:1}			
			\STATE
			Compute the minimum number of samples $N_i$ according to \eqref{Min_Data}
			
			\STATE
			Solve the SOP$_\varsigma$ in \eqref{SOP_2} with acquired $\hat{N}_i$ and $N_i$, and obtain $\eta^*_{\varsigma_i}$ and $\Theta^*_i$
			\ENSURE
			if $\eta^*_{\varsigma_i} + \varepsilon_{1_i} \leq 0$, then $\Theta^*_i$ is a feasible solution for ROP with a confidence of at least $1-\beta_{1_i}-\beta_{2_i}$
		\end{algorithmic}
	\end{center}
\end{algorithm}

In order to compute the required number of samples in Theorem~\ref{Thm:2}, one needs to first compute $\mathscr{L}_{g_{i_k}}$. In the Appendix, we propose a systematic approach to compute $\mathscr{L}_{g_{i_k}}$ for both linear and nonlinear agents. 

\begin{remark}
	Note that the Lipschitz constant of dynamics and an upper bound on unknown dynamics are the minimal information we need from the system in our data-driven setting. However, one can utilize the proposed approach in~\cite{wood1996estimation} to estimate the Lipschitz constant of dynamics via sampled data. As for an upper bound on unknown dynamics, it can be computed based on the range of the state set.
\end{remark}

{\bf Case Study: Vehicle Platooning (continued).} We assume the model of each agent is unknown. The main goal is to compositionally construct a BC for the multi-agent AV based on SBCs of individual agents via data collected from trajectories of agents and solving $\text{SCP}_\varsigma$~\eqref{SOP_2}. Accordingly, we formally estimate the collision risk for the multi-agent AV within a finite time horizon with some desirable confidence.

We first fix the structure of our sub-barrier certificates as $\mathcal B_i(q_i,x_i) = q_{i_1}d_i^2 + q_{i_2} d_i^4 + q_{i_3} d_i\nu_i + q_{i_4}\nu_i^4 + q_{i_5}\nu_i^2$, for all  $i\in\{1,\ldots,100\}$. According to Algorithm~\ref{Alg:1}, we fix the threshold $\varepsilon_{1_i} = 0.08$ and the confidence $\beta_{2_i} = 10^{-4}$, a-priori. In order to reduce the number of decision variables, and accordingly, reduce the number of data in~\eqref{Min_Data} required for solving the $\text{SOP}_{\varsigma}$~\eqref{SOP_2}, we a-priori fix $\lambda_i = 10, \psi_i = 10^{-4}, \alpha_i = 10^{-4}, \rho_i = 9\times 10^{-7},$ for all  $i\in\{1,\ldots,100\}$. Now we need to compute $\mathscr L_g$ which is required for computing the minimum number of data. We construct matrix $P_i$ based on coefficients of the SBC as 
\begin{align*}
P_i = \begin{bmatrix}
q_{i_1} & 0 & \frac{q_{i_3}}{2} & 0\\
0 & q_{i_2} & 0 & 0 \\
\frac{q_{i_3}}{2} & 0 & q_{i_5} &  0 \\
0 & 0 & 0 & q_{i_4}
\end{bmatrix}\!\!.
\end{align*}
We enforce $-0.001 \leq q_{i_1},q_{i_2},q_{i_3},q_{i_5}\leq 0.001, -0.14 \leq q_{i_4}\leq 0.14$ as discussed in Remark~\ref{Gerschgorin}. We assume that $\kappa_i \in \{0.9,0.99\}$ with the cardinality $m_i = 2$. Then according to Lemma~\ref{Lemma:2}, we compute $\mathscr L_{g_{i_k}} = 1.7804$, $\ k\in\{1,\dots,m_i\}$, and accordingly, $\varepsilon_{2_{i_k}}=(\frac{\varepsilon_{1_i}}{\mathscr{L}_{g_{i_k}}})^{n_i+p_i} = (\frac{0.08}{1.7804})^{3} = 9.0723\times 10^{-5}$.  

\begin{figure}[h]
	\centering
	\includegraphics[scale=0.17]{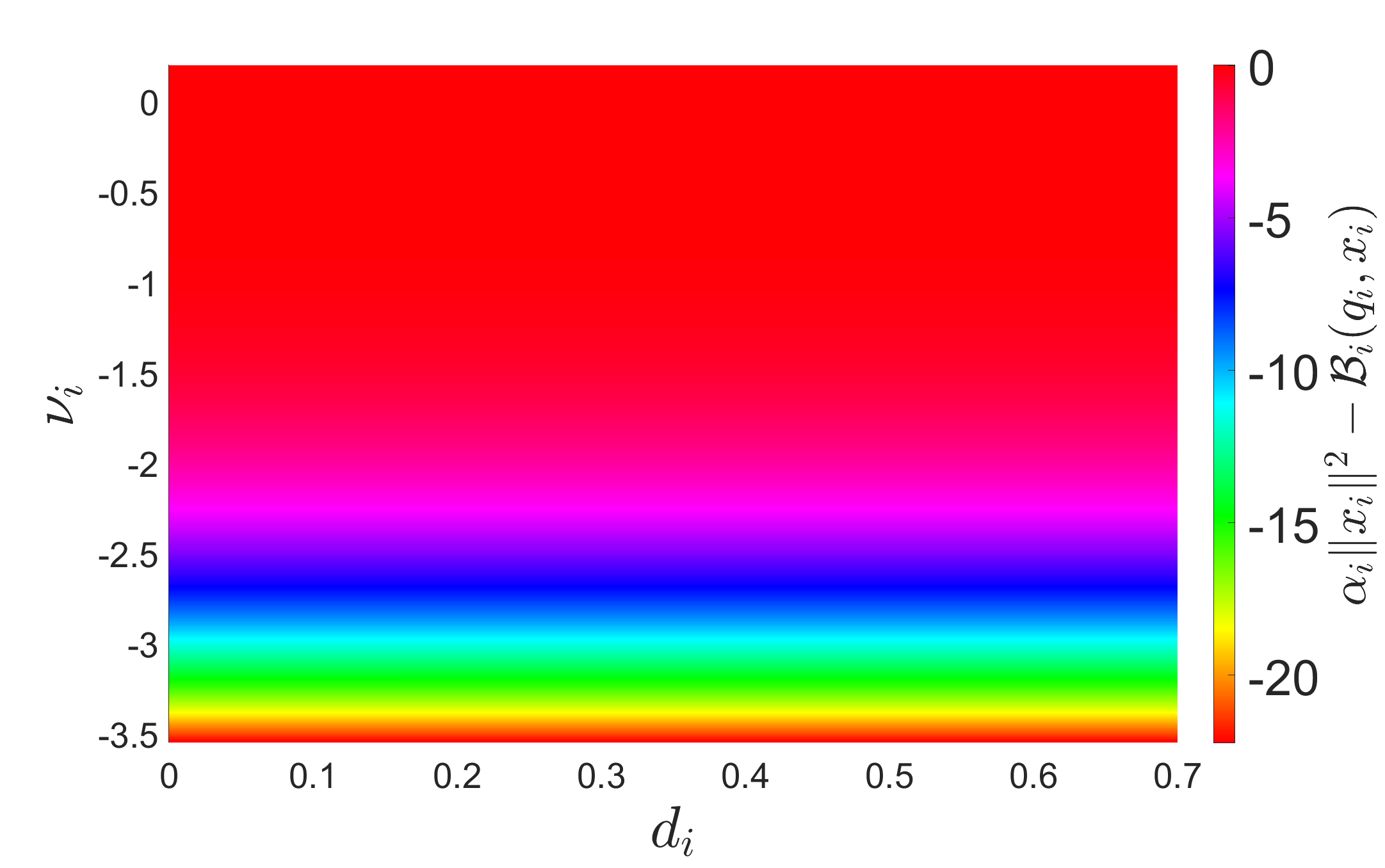}\hspace{0.5cm}
	\includegraphics[scale=0.17]{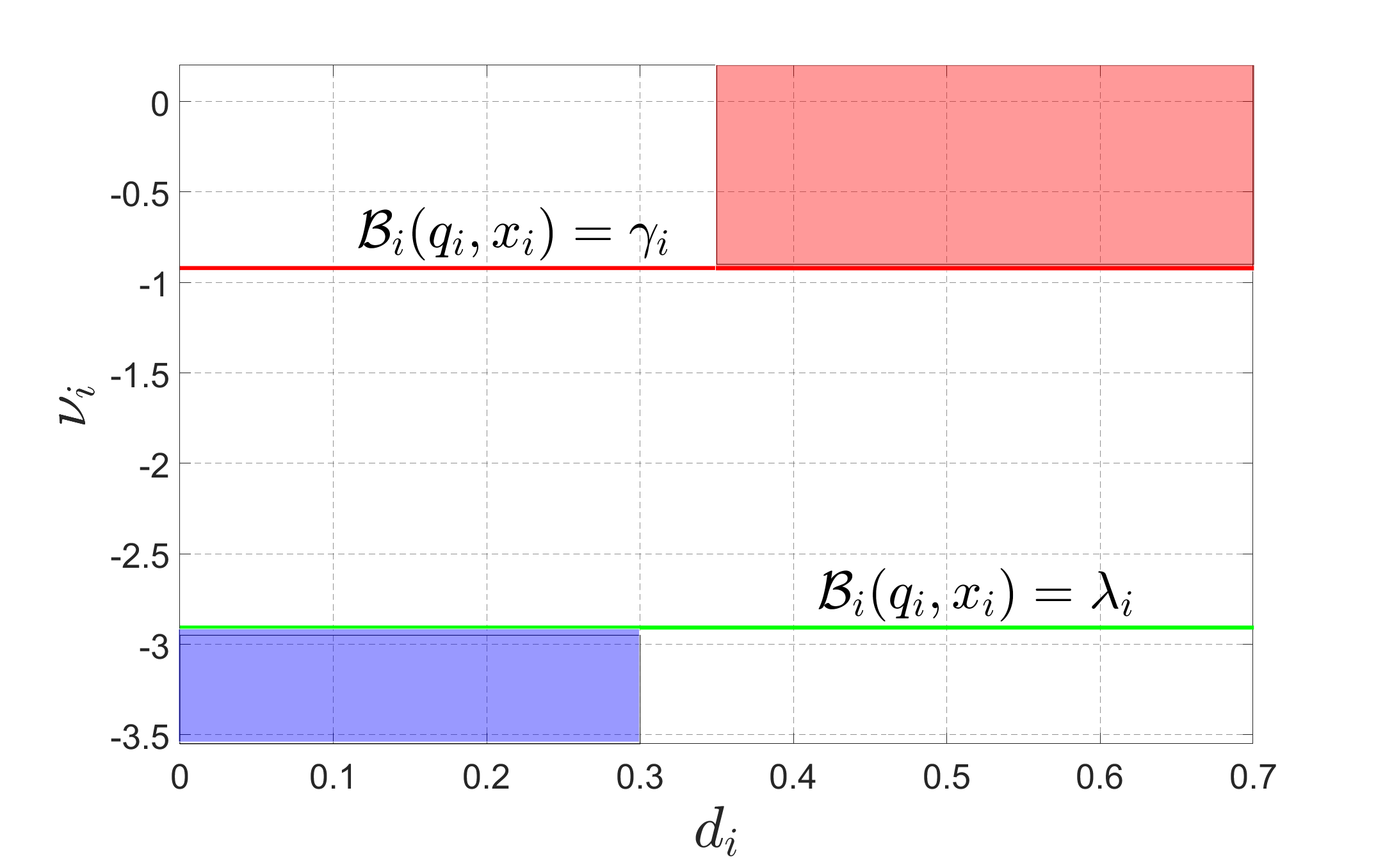}
	\caption{Left: Satisfaction of condition~\eqref{Eq:8_1}. As observed, this condition is negative for all ranges of $d_i \in [0,0.7]$ and $\nu_i \in [-3.55,0.2]$. Right: Satisfaction of conditions~\eqref{Eq:8_2}-\eqref{Eq:8_3} based on SBC of each unknown vehicle. Pink and purple boxes are initial and collision sets, respectively.}
	\label{Fig3}
\end{figure}

Now we have all the required ingredients to compute $N_i$. The minimum number of data required for solving $\text{SOP}_{\varsigma}$ in~\eqref{SOP_2} is computed as $N_i = 244993.$
Now we need to compute $\hat N_i$ which is required for solving the $\text{SOP}_{\varsigma}$~\eqref{SOP_2} (\emph{i.e.,} condition $\bar g_{i_5}$). According to Lemma~\ref{Lemma:1}, we fix $\mu_i = 0.08$ with a-priori confidence $\beta_{1_i} = 10^{-4}$ and compute $\hat N_i = 11$. We refer the interested reader to~\cite{Ali_ADHS21} for more details on the quantification of $\hat N_i$. By leveraging the computed parameters, the $\text{SOP}_{\varsigma}$ in~\eqref{SOP_2} is solved with $\kappa_i = 0.99$ and the following decision variables:
\begin{align}\notag
\mathcal B_i(q_i,x_i) &= -0.0008 d_i^2 + 0.001 d_i^4 + 0.0001 d_i\nu_i + 0.14\nu_i^4 - 0.0006\nu_i^2, \\\label{decision}
\eta_{\varsigma_i}^*&=-0.085, ~\gamma^*_i = 0.1.
\end{align}
Since $\eta_{\varsigma_i}^* + \varepsilon_{1_i} = -0.005 \leq 0$, according to Theorem~\ref{Thm:2}, one can guarantee that the constructed SBC via collected data together with other decision variables in~\eqref{decision} are valid for the original  ROP~\eqref{ROP} with a confidence of at least $1-\beta_{1_i} - \beta_{2_i}  = 1 - 10^{-4} - 10^{-4}$. Satisfaction of conditions~\eqref{Eq:8_1}-\eqref{Eq:8_3} via constructed SBC from data is illustrated in Fig.~\ref{Fig3}.

\begin{remark}
	Note that the sample space of $\hat w_i$ is defined with respect to the interconnection topology according to~\eqref{Eq:4}. In particular, since each vehicle is only connected to its proceeding vehicle in a cascade interconnection topology, the sample of $\hat w_i$ for each agent contains only the state of the proceeding vehicle.  
\end{remark}

In the next section, we consider the multi-agent AV $\mathcal A$ as in Definition~\ref{Def:2} and provide a compositional framework for the construction of BC for $\mathcal A$ using SBC of $\mathcal A_i$.

\section{Compositionality Results for Multi-Agent AVs}\label{Sec:Com}

In this section, we analyze multi-agent AV $\mathcal A = \mathcal{I}(\mathcal A_1,\ldots,\mathcal A_M)$ by driving a small-gain type compositional condition
and discussing how to construct a BC for the multi-agent AV based on SBC of its individual agents. The construed BC is useful to estimate the collision risk for multi-agent AVs according to Theorem~\ref{Thm:1}. Before presenting the main compositionality result of the work, we define $ \Lambda\Let\mathsf{diag}(\hat\lambda_1,\ldots,\hat\lambda_M)$ with  $\hat\lambda_i = 1 - \kappa_i$, and  $\Delta\Let\{\hat\delta_{ij}\}$ with $\hat\delta_{ij} = \frac{\rho_i}{\alpha_j}$, where $\hat\delta_{ii}=0$, $\forall i\in\{1,\cdots,M\}$.

In the next theorem, we show that how one can construct a BC for the multi-agent AV $\mathcal A$ using SBC of $\mathcal A_i$.

\begin{theorem}\label{Thm:3}
	Consider the multi-agent AV
	$\mathcal A=\mathcal{I}(\mathcal A_1,\ldots,\mathcal A_M)$ induced by $M\in\mathbb N_{\geq1}$ agents~$\mathcal A_i$. Suppose that each $\mathcal A_i$ admits an SBC $\mathcal B_i$ with a confidence of at least $1-\beta_{1_i}-\beta_{2_i}$, as proposed in Theorem~\ref{Thm:2}. If 
	\begin{align}\label{Eq:43}
	&\sum_{i=1}^{M} \lambda_i > \sum_{i=1}^{M} \gamma_i,\\\label{Eq:44}
	\mathds{1}_M^\top(-\Lambda+\Delta) =: \big[\pi_1;\dots;\pi_M\big]^\top& <  0,~\text{equivalently}, ~ \pi_i < 0,~ \forall i\in\{1,\dots,M\},
	\end{align} 
	then $$\mathcal B(q,x) :=\ \sum_{i=1}^{M}\mathcal B_i(q_i,x_i)$$ is a BC for the multi-agent AV $\mathcal A$ with a confidence of at least $1-\sum_{i=1}^{M}\beta_{1_i}-\sum_{i=1}^{M}\beta_{2_i}$, and with
	\begin{align*}
	&\gamma := \sum_{i=1}^{M}\gamma_i, \quad \lambda := \sum_{i=1}^{M} \lambda_i, \quad \psi := \sum_{i=1}^{M}\psi_i, \\
	&\kappa:= 1 + \pi, ~ \text{with} ~  \max_{1\leq i\leq M} \pi_i < \pi < 0, ~ \text{and} ~ \pi \in (-1,0).
	\end{align*}
\end{theorem}

\begin{remark}
	Note that $\hat\lambda_i$ and $\hat\delta_{ij}$ are used to capture, respectively, the gain of each individual agent and its interaction gain with other agents in the interconnection topology, \emph{i.e.,} $\kappa_{i}, \rho_{i}$. Those $\hat\lambda_i$ and $\hat\delta_{ij}$ are then utilized for the construction of $\Lambda$ and $\Delta$, and accordingly, establishing the compositionality condition in~\eqref{Eq:44}. 
\end{remark}

\begin{remark}
	If one is interested in enforcing the compositionality conditions during solving $\text{SOP}_\varsigma$ in~\eqref{SOP_2}, enforcing $\mathds{1}_M^\top(-\Lambda+\Delta)< 0$ is not directly possible due to having a bilinearity between decision variables $\rho_i$ and $\alpha_j$. As a potential solution, one can a-priori fix either $\rho_i$ or $\alpha_j$ to resolve the bilinearity $\frac{\rho_i}{\alpha_j}$ and then enforce compositionality conditions~\eqref{Eq:43}-\eqref{Eq:44} as two additional conditions while solving $\text{SOP}_\varsigma$ in~\eqref{SOP_2}.
\end{remark}

{\bf Case Study: Vehicle Platooning (continued).} We now proceed with Theorem~\ref{Thm:3} to construct a BC for the multi-agent AV with a level of confidence using SBCs of individual agents constructed from data. By constructing $\Lambda\Let\mathsf{diag}(\hat\lambda_1,\ldots,\hat\lambda_M)$ with  $\hat\lambda_i = 0.01$, and  $\Delta\Let\{\hat\delta_{ij}\}$ with $\hat\delta_{ij} = \frac{\rho_i}{\alpha_j} = 9 \times 10^{-3}$, one can readily verify that the compositionality condition $\mathds{1}_M^\top(-\Lambda+\Delta) < 0$ is satisfied. Moreover, the compositionality condition~\eqref{Eq:43} is also met since $\lambda_i > \gamma_i,$ for all $i\in \{1,\dots,100\}$. Then by employing the results of Theorem~\ref{Thm:3}, one can conclude that $\mathcal  B(q, x) = \sum_{i=1}^{100} (-0.0008 d_i^2 + 0.001 d_i^4 + 0.0001 d_i\nu_i + 0.14\nu_i^4 - 0.0006\nu_i^2)$ is a BC for the multi-agent AV with $\gamma = 10, \lambda = 1000, \kappa= 0.99$ and $\psi = 0.01$, with a confidence of at least $1-\sum_{i=1}^{100}\beta_{1_i}-\sum_{i=1}^{100}\beta_{2_i} = 98\%$.

\begin{figure}[h]
	\centering 
	\includegraphics[width=0.42\linewidth]{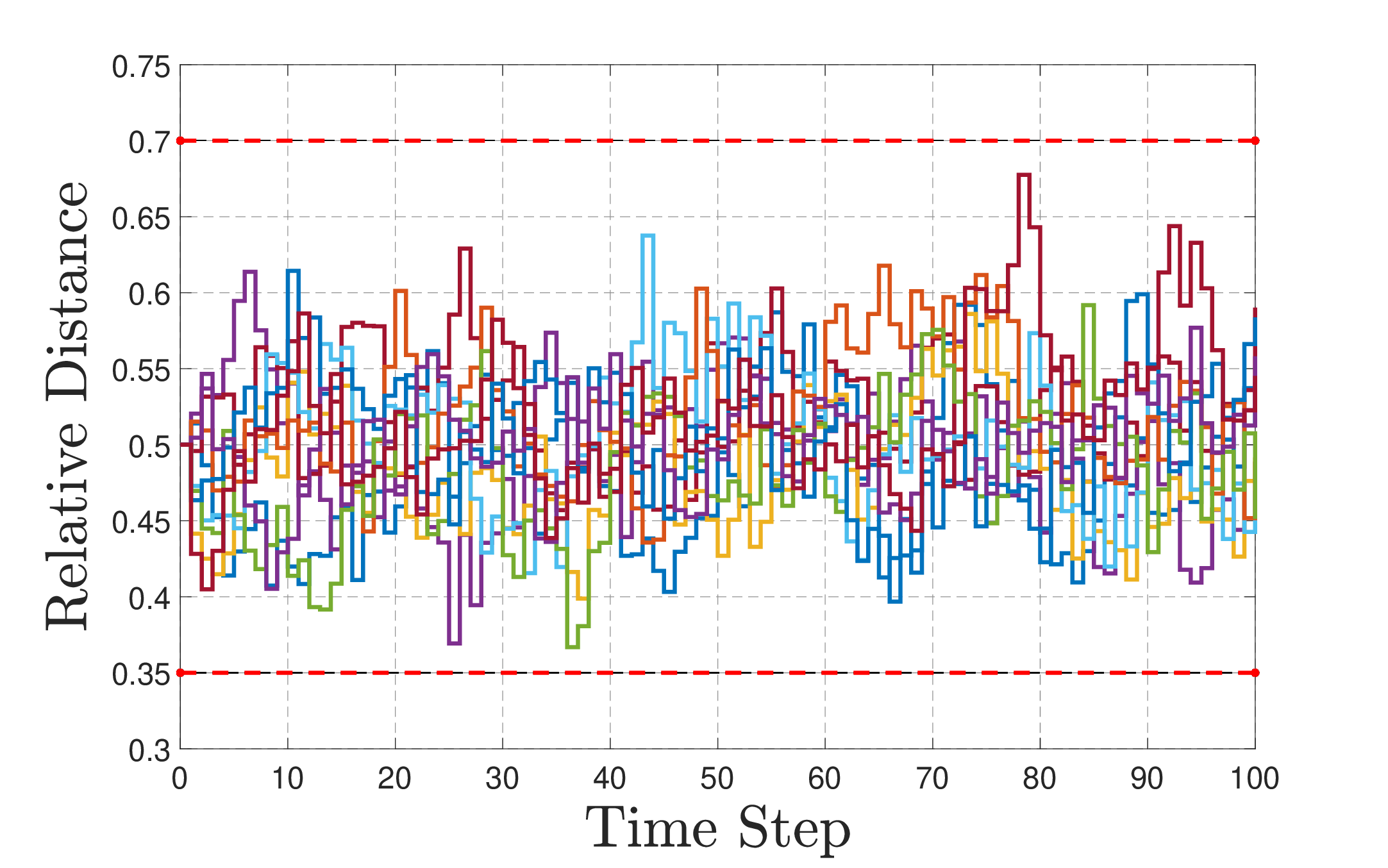}\hspace{0.4cm}
	\includegraphics[width=0.42\linewidth]{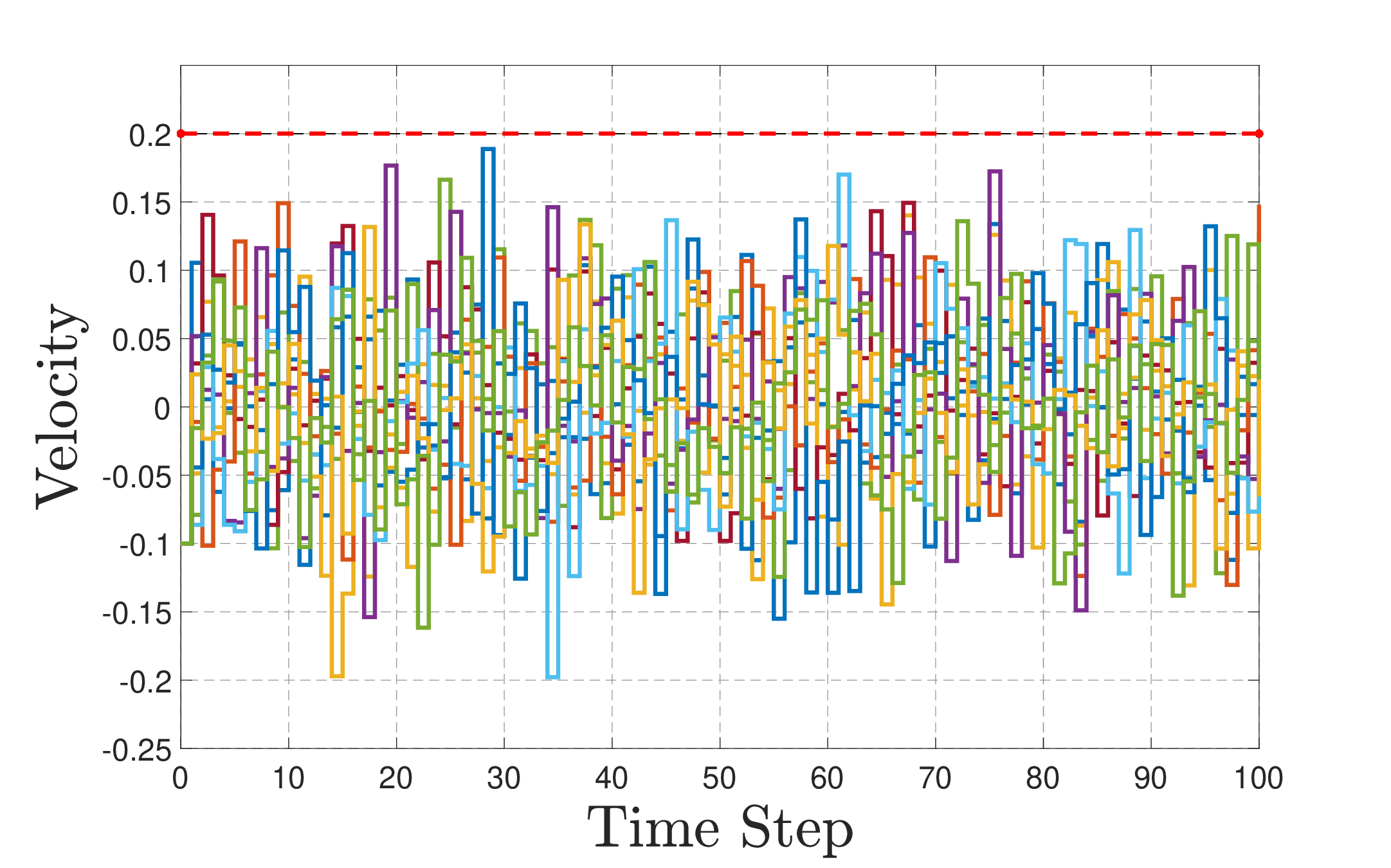}
	\caption{Closed-loop (relative distance and velocity) trajectories of a representative vehicle with $10$ different noise realizations in a network of
		$100$ vehicles.}
	\label{Fig4}
\end{figure}

By employing Theorem~\ref{Thm:1}, we guarantee that the collision risk for the multi-agent AV is at most $1\%$ with a confidence of at least $98\%$ during the time horizon $\mathcal T = 100$, \emph{i.e.,}
\begin{equation}\label{threshold}
\PP^{N}\Big\{\PP\big\{\mathcal A\vDash_{100} \varphi\big\}\leq 0.01\Big\}\geq 0.98.
\end{equation}
In order to verify our results, we assume that we have access to the model of agents and plot the closed-loop state (relative distance and velocity) trajectories of a representative vehicle with $10$ different noise realizations as in Fig.~\ref{Fig4}. As it can be observed, none of $10$ state trajectories violates the safety specification, which is in accordance with our collision risk guarantee in~\eqref{threshold}.

\noindent The SBC computation for each vehicle takes $12$ minutes on a Windows machine (Intel i7-8665U CPU with 32 GB of RAM).

In case the compositionality conditions~\eqref{Eq:43}-\eqref{Eq:44} are not satisfied, we propose, in the next section, an alternative compositional approach that does not require any compositionality conditions but at the cost of providing a potentially conservative collision risk for the multi-agent AVs.

\section{Relaxed Compositional  Approach}\label{Sec: Com_Relaxed}
In this section, we propose an alternative compositional approach which is relax compared to the small-gain reasoning in the sense that (i) condition~\eqref{Eq:8_1} is no longer needed, (ii) compositionality conditions~\eqref{Eq:43}-\eqref{Eq:44} are no more required, (iii) $\kappa_i$ in condition~\eqref{Eq:8_4} is equal to one, and (iv) the required number of data in Theorem~\ref{Thm:2} is reduced since $m_i = 1$. On the downside, the provided collision risk for the multi-agent AV is potentially conservative. In our proposed alternative setting,  we omit $g_{i_2}$ and modify some other conditions of ROP in~\eqref{ROP} as
\begin{align}\notag
g_{i_5}&=\EE\Big[\mathcal{B}_i(q_i,f_i(x_i,w_i,\varsigma_i))\mid x_i, w_i\Big]-\mathcal{B}_i(q_i,x_i)-\rho_i\Vert w_i\Vert^2-\psi_i, \\\label{Eq:101}
g_{i_6} &= \gamma_i - \lambda_i - \varrho_i,
\end{align}
for some $\varrho_i < 0$. We now compute the collision risk for \emph{each individual agent} $\mathcal A_i$ via the following theorem.

\begin{theorem}\label{Thm:4}
	Consider an agent $\mathcal A_i$ defined in Definition~\ref{Def:1} and a finite time horizon $\mathcal T\in \mathbb{N}$. Suppose that there exists a non-negative SBC $\mathcal B_i$ satisfying conditions \eqref{Eq:8_2}-\eqref{Eq:8_4} with $\lambda_i > \gamma_i$. Then the collision risk for each agent $\mathcal A_i$ is computed as
	\begin{equation}\label{Eq:12}
	\PP\Big\{\mathcal A_i\models_{\mathcal T} \varphi_i\Big\}\leq\delta_i,~\text{with}~ \delta_i\!=  \frac{\gamma_i + \hat\psi_i\mathcal T}{\lambda_i},
	\end{equation}
	where $\hat\psi_i = \rho_i \Vert w_i\Vert^2_\infty + \psi_i$.
\end{theorem}

In comparison with the small-gain compositional approach, one needs here to solve SOP$_\varsigma$ in~\eqref{SOP_2} without $g_{i_2}$ and with $g_{i_5}, g_{i_6}$ as in~\eqref{Eq:101}. Moreover, the interconnection of  $\mathcal A_i$, $\forall i\in \{1,\ldots,M\}$, is defined here by an interconnection map $h:\prod_{i=1}^{M}X_i\to\prod_{i=1}^{M}W_i$, and accordingly, the interconnection constraint in~\eqref{Eq:4} can be generalized to $[w_1;\ldots;w_M]=h(x_1,\dots,x_M)$. Similar to Theorem~\ref{Thm:2}, one can relate the optimal value of $\text{SOP}_\varsigma$ to that of ROP in~\eqref{ROP} with the new conditions in~\eqref{Eq:101}, and consequently, formally quantify collision risks of unknown agents $\mathcal A_i$ based on the number of data as in~\eqref{Min_Data} with $m_i = 1$, and a required level of confidence. We now propose the relaxed compositionality results of this section.
\begin{theorem}\label{Thm:5}
	Consider the multi-agent AV $\mathcal A = (X,\varsigma,f)$ composed of $M$ agents $\mathcal A_i=(X_i,W_i,\varsigma_i,f_i)$, with a collision event $\varphi = \varphi_1 \times \varphi_2 \times \dots \times \varphi_M$.
	Let the collision risk for each $\mathcal A_i$ be at most $\delta_i$ (in the sense of Theorem~\ref{Thm:4}) with a confidence of at least $1-\beta_i$, with $\beta_i = \beta_{1_i}+\beta_{2_i}, \forall i\in \{1,\dots,M\}$, i.e., $$\PP\Big\{\PP\big\{\mathcal A_i\models \varphi_i \big\}\leq  \delta_i\Big\} \ge 1 - \beta_i.$$ Then the collision risk for the multi-agent AV $\mathcal A$ is at most $\sum_{i =1}^M \delta_i$ with a confidence of at least $1-\sum_{i =1}^M\beta_i$, \emph{i.e.,} $$\PP\Big\{\PP\big\{\mathcal A_1\models \varphi_1 \cup\dots\cup\mathcal A_M\models\varphi_M\big\}\leq \sum_{i =1}^M \delta_i\Big\} \ge 1-\sum_{i =1}^M\beta_i.$$
\end{theorem}

\begin{remark}
	As it can be observed, the confidence level for the both small-gain and relaxed compositional approaches is the same, \emph{i.e.,} $1-\sum_{i=1}^{M}\beta_{1_i}-\sum_{i=1}^{M}\beta_{2_i}$. In the relaxed compositional approach, the collision risk for the multi-agent AV is computed based on the linear combination of estimated collision risks for individual agents, \emph{i.e.}, $\sum_{i=1}^{M}\delta_{i}$. In contrast, using the small-gain compositional approach, we first construct the overall BC based on SBC of agents and then estimate the collision risk of the multi-agent AV via results of Theorem~\ref{Thm:1}. Accordingly, the estimated collision risk via the small-gain compositional approach is potentially less conservative but at the cost of satisfying some additional conditions.
\end{remark}

In the next section, we present our data-driven approaches for \emph{deterministic} multi-agents AVs without any stochasticity $\varsigma_i$.  

\section{Deterministic (Multi)Agent AVs}\label{Sec: Deterministic}

In this section, we consider dynamics of each agent as a discrete-time \emph{deterministic} system given by
\begin{equation}\label{Eq:13}
\mathcal A_i\!:x_i(k+1)=f_i(x_i(k),w_i(k)), \quad k\in\mathbb N,
\end{equation}
with $f_i:X_i\times W_i \rightarrow X_i$, and represent it with $\mathcal A_i=(X_i,W_i,f_i)$. The multi-agent AV $\mathcal A$ without interactions $w_i$, constructed as a composition of several agents $\mathcal A_i$ with interactions, can be represented by $\mathcal A=(X,f)$ with $f: X \rightarrow X$, and  $\mathcal A\!:x(k+1)=f(x(k))$. We now define barrier certificates for \emph{deterministic} multi-agent AVs as the next definition.

\begin{definition}\label{Def:6}
	Given the deterministic multi-agent AV $\mathcal A = (X,f)$ with initial and collision sets $X_{0}, X_{c}\subseteq X$, a function $\mathcal{B}\!:X\rightarrow\mathbb{R}$ is called a barrier certificate (BC) for $\mathcal A$ if there exist $\gamma, \lambda \in\mathbb R$, with $\lambda > \gamma$, and $0<\kappa <1$, such that conditions~\eqref{Eq:6_1}-\eqref{Eq:6_2} are satisfied, and
	\begin{align}\label{Eq:6_4}
	\mathcal{B}(f(x))\leq\kappa\mathcal{B}(x),\quad\quad\forall x \in X.
	\end{align}
\end{definition} 

Now we employ Definition~\ref{Def:6} and quantify the collision risk for multi-agent AVs via the next theorem.

\begin{theorem}\label{Thm:9}
	Consider a deterministic multi-agent AV $\mathcal A = (X,f)$. Suppose $\mathcal B$ is a BC for $\mathcal A$ as in Definition~\ref{Def:6}. Then the collision risk is zero for the multi-agent AV within an \emph{infinite time horizon}, i.e., $x_{x_{0}}(k) \cap X_{c} = \emptyset$ for any $x_{0}\in X_{0}$ and any $k\in \mathbb N$.
\end{theorem}

Since searching for the BC as in Definition~\ref{Def:6} is computationally expensive, we now define sub-barrier certificates for deterministic agents $\mathcal A_i$ as the following definition.
\begin{definition}\label{Def:5_1}
	Consider a deterministic agent $\mathcal A_i = (X_i,W_i,f_i)$, and $X_{0_i}, X_{c_i}\subseteq X_i$ as its initial and collision sets, respectively. A function $\mathcal{B}_i:X_i\rightarrow\mathbb{R}$ is called a sub-barrier certificate (SBC) for $\mathcal A_i$ if there exist $\gamma_i, \lambda_i \in\mathbb R$, $\alpha_i,\rho_i\in\mathbb R_0^+$, and $0<\kappa_i <1$, such that conditions~\eqref{Eq:8_1}-\eqref{Eq:8_3} are satisfied, and $\forall x_i \in X_i, \forall w_i \in W_i$,
	\begin{align}\label{Eq:14_1}
	\mathcal{B}_i(f_i(x_i,w_i))\leq\kappa_i\mathcal{B}_i(x_i)+\rho_i\Vert w_i\Vert^2.
	\end{align}
\end{definition}

Here, we assume that the transition map $f_i$ in~\eqref{Eq:13} is unknown.
We similarly recast the conditions of SBC as the proposed ROP in~\eqref{ROP}, but without $g_{i_1}$, and with
\begin{align}\label{new}
&g_{i_5}=\mathcal{B}_i(q_i,f_i(x_i,w_i))-\kappa_i\mathcal{B}_i(q_i,x_i) -\rho_i\Vert w_i\Vert^2.
\end{align}
Since the proposed ROP in~\eqref{ROP} has infinitely many constraints and a precise transition map $f_i$ is also needed for solving the problem, we employ the proposed $\text{SOP}_\text{N}$ in~\eqref{SOP_1} instead of solving the ROP in~\eqref{ROP}. Note that we do not need to employ $\text{SOP}_\varsigma$ in~\eqref{SOP_2} since there is no stochasticity inside the model. Similar to Theorem~\ref{Thm:2}, we propose the next theorem to relate the optimal value of $\text{SOP}_\text{N}$ in~\eqref{SOP_1} to that of ROP in~\eqref{ROP}, and accordingly, formally quantify the SBC of unknown agents $\mathcal A_i$ based on the number of data and a required level of confidence. 

\begin{theorem}\label{Thm:7_1}
	Consider the unknown \emph{deterministic} agent $\mathcal A_i$ as in~\eqref{Eq:13}, and initial and collision regions $X_{0_i}$ and $X_{c_i}$, respectively. Let $g_{i_2}$-$g_{i_4}$ be Lipschitz continuous with respect to $x_i$, and $g_{i_5}$ as in~\eqref{new} be Lipschitz continuous with respect to $x_i,w_i$, with Lipschitz constants $\mathscr{L}_{g_{i_2}}$-$\mathscr{L}_{g_{i_4}}$, $\mathscr{L}_{g_{i_{5_k}}}$, respectively. Consider the corresponding $\text{SOP}_\text{N}$ in~\eqref{SOP_1} with its associated optimal value $\eta^*_{N_i}$ and solution $\Theta^*_i = [\gamma^*_i;\lambda^*_i;\alpha^*_i;\rho^*_i;{q}^*_{i_1};\dots;q^*_{i_r}]$, with $N_i\geq N_i\big(\bar \varepsilon_{2_i},\beta_{2_i}\big)$, as in~\eqref{Min_Data} with $\varepsilon_{1_i},\beta_{2_i} \in [0,1]$  where $\varepsilon_{1_i} \leq \mathscr{L}_{g_{i_k}} = \max \big\{\mathscr{L}_{g_{i_2}},\mathscr{L}_{g_{i_3}},\mathscr{L}_{g_{i_4}},\mathscr{L}_{g_{i_{5_k}}}\big\}$. If $\eta^*_{N_i} + \varepsilon_{1_i} \leq 0$, then the solution $\Theta^*_i$ is a feasible solution for ROP in~\eqref{ROP} with a confidence of at least $1-\beta_{2_i}$.
\end{theorem}

Proof of Theorem~\ref{Thm:7_1} is similar to that of Theorem~\ref{Thm:2} and is omitted here. 

\subsection{Small-Gain Compositional Approach} 
Given a deterministic multi-agent AV $\mathcal A = \mathcal{I}(\mathcal A_1,\ldots,\mathcal A_M)$, we show in the next theorem that how to construct a BC for the multi-agent AV $\mathcal A$ using SBCs of $\mathcal A_i$.

\begin{theorem}\label{Thm:3_1}
	Consider the deterministic multi-agent AV
	$\mathcal A=\mathcal{I}(\mathcal A_1,\ldots,\mathcal A_M)$ induced by $M\in\mathbb N_{\geq1}$ agents~$\mathcal A_i$ as in~\eqref{Eq:13}. Suppose that each $\mathcal A_i$ admits an SBC $\mathcal B_i$ with a confidence of at least $1-\beta_{2_i}$, as proposed in Theorem~\ref{Thm:7_1}. If the compositional conditions~\eqref{Eq:43}-\eqref{Eq:44} are satisfied, then $$\mathcal B(q,x) := \sum_{i=1}^{M}\mathcal B_i(q_i,x_i)$$ is a BC for the multi-agent AV $\mathcal A$ with a confidence of at least $1-\sum_{i=1}^{M}\beta_{2_i}$, and with
	\begin{align*}
	&\gamma := \sum_{i=1}^{M}\gamma_i, \quad \lambda := \sum_{i=1}^{M} \lambda_i, \quad \kappa:= 1 + \pi, ~\text{with} ~  \max_{1\leq i\leq M} \pi_i < \pi < 0, ~ \text{and} ~ \pi \in (-1,0).
	\end{align*}
\end{theorem}

Proof of Theorem~\ref{Thm:3_1} is similar to that of Theorem~\ref{Thm:3} and is omitted here.

\subsection{Relaxed Compositional Approach} 
In the case that the compositionality conditions are not satisfied, we propose a relax compositional approach, in which (i) condition~\eqref{Eq:8_1} is no longer needed, (ii) compositionality conditions~\eqref{Eq:43}-\eqref{Eq:44} are no more required, (iii) $\kappa_i$ in condition~\eqref{Eq:14_1} is equal to one,  and (iv) the required number of data in Theorem~\ref{Thm:7_1} is reduced, but at the cost of providing the collision risk estimation for multi-agent AVs in \emph{finite time horizons}. To do so, we first propose a new definition of SBC for deterministic agents $\mathcal A_i$ as the following.
\begin{definition}\label{Def:5}
	Consider the deterministic agent $\mathcal A_i = (X_i,W_i,f_i)$, and $X_{0_i}, X_{c_i}\subseteq X_i$ as its initial and collision sets, respectively. A function $\mathcal{B}_i\!:X_i\rightarrow\mathbb{R}$ is called a sub-barrier certificate (SBC) for $\mathcal A_i$ if there exist $\rho_i\in\mathbb R_0^+$, $\gamma_i, \lambda_i \in\mathbb R$ with $\gamma_i + \rho_i\Vert w_i\Vert^2_\infty\mathcal T_i < \lambda_i$, within a finite time horizon $\mathcal T_i\in \mathbb{N}$, such that conditions~\eqref{Eq:8_2}-\eqref{Eq:8_3} are satisfied, and $\forall x_i \in X_i, \forall w_i \in W_i$,
	\begin{align}\label{Eq:14}
	\mathcal{B}_i(f_i(x_i,w_i))\leq\mathcal{B}_i(x_i)+\rho_i\Vert w_i\Vert^2.
	\end{align}
\end{definition}

We now employ Definition~\ref{Def:5} and estimate the collision risk for each agent $\mathcal A_i$ in a \emph{finite time horizon} via the next theorem.

\begin{theorem}\label{Thm:6}
	Consider the deterministic agent $\mathcal A_i$ in~\eqref{Eq:13} and a finite time horizon $\mathcal T_i\in \mathbb{N}$. Suppose $\mathcal B_i$ is an SBC for $\mathcal A_i$ as in Definition~\ref{Def:5}. Then, the collision risk is zero for each agent $\mathcal A_i$ within the \emph{finite time horizon} $\mathcal T_i = \frac{\lambda_i - \gamma_i}{\rho_i\Vert w_i\Vert^2_\infty}$, \emph{i.e.,} $x_{x_{0_i}}(k) \cap X_{c_i} = \emptyset$ for any $x_{0_i}\in X_{0_i}$ and any $k\in [0 , \mathcal T_i)$, with $\mathcal T_i = \frac{\lambda_i - \gamma_i}{\rho_i\Vert w_i\Vert^2_\infty}$.
\end{theorem}

We similarly recast the conditions of SBC as the proposed ROP in~\eqref{ROP}, but without $g_{i_1},g_{i_2}$, and with
\begin{align*}
&g_{i_5}(x_i,\Theta_i)=\mathcal{B}_i(q_i,f_i(x_i,w_i))-\mathcal{B}_i(q_i,x_i) -\rho_i\Vert w_i\Vert^2,\\
&g_{i_6}(x_i,\Theta_i)= \gamma_i + \rho_i\Vert w_i\Vert^2_\infty\mathcal T - \lambda_i - \varrho_i, 
\end{align*}
for some $\varrho_i <0$. Similar to Theorem~\ref{Thm:7_1}, one can relate the optimal value of $\text{SOP}_\text{N}$ in~\eqref{SOP_1} to that of ROP in~\eqref{ROP}, and accordingly, quantify the SBC of unknown agents $\mathcal A_i$ based on the number of data as in~\eqref{Min_Data} with $m_i = 1$, and a required level of confidence. We now propose our relaxed compositional approach for multi-agent AVs $\mathcal A$ with unknown \emph{deterministic} dynamics via the next theorem.
\begin{theorem}\label{Thm:8}
	Consider the deterministic multi-agent AV $\mathcal A = (X,f)$ composed of $M$ agents $\mathcal A_i=(X_i,W_i,f_i)$, $i\in \{1,\dots,M\}$, with a collision event $\varphi = \varphi_1 \times \varphi_2 \times \dots \times \varphi_M$.
	Let the collision risk for each $\mathcal A_i$ be zero (in the sense of Theorem~\ref{Thm:6}) for a finite time horizon $\mathcal T_i = \frac{\lambda_i - \gamma_i}{\rho_i\Vert w_i\Vert^2_\infty}$ with a confidence of at least $1-\beta_{2_i},\forall i\in \{1,\dots,M\}$, \emph{i.e.,} $\PP\Big\{\mathcal A_i\nvDash\varphi_i \Big\} \ge 1 - \beta_{2_i}$. Then the collision risk for the multi-agent AV $\mathcal A$ is zero within the \emph{finite time horizon} $\mathcal T = \min_{i\in\{1,\dots,M\}}\{\mathcal T_i\}$, with a confidence of at least $1-\sum_{i =1}^M\beta_{2_i}$, \emph{i.e.,} \begin{align*}
	\PP\Big\{\mathcal A_1\nvDash\varphi_1 \cap\dots\cap\mathcal A_M\nvDash\varphi_M\Big\} \ge 1-\sum_{i =1}^M\beta_{2_i}.
	\end{align*}
\end{theorem}

Proof of Theorem~\ref{Thm:8} is similar to that of Theorem~\ref{Thm:5} and is omitted here.

The confidence level for multi-agent AVs with unknown \emph{deterministic} dynamics in the both small-gain and relaxed compositional approaches  is the same, \emph{i.e.,} $1-\sum_{i=1}^{M}\beta_{2_i}$. However, the collision risk in the relaxed compositional approach is estimated in \emph{finite time} horizons, whereas small-gain compositional approach provides the collision risk estimation in \emph{infinite time} horizons but at the cost of fulfilling some additional conditions.

\begin{figure}
	\centering
	\includegraphics[scale=0.24]{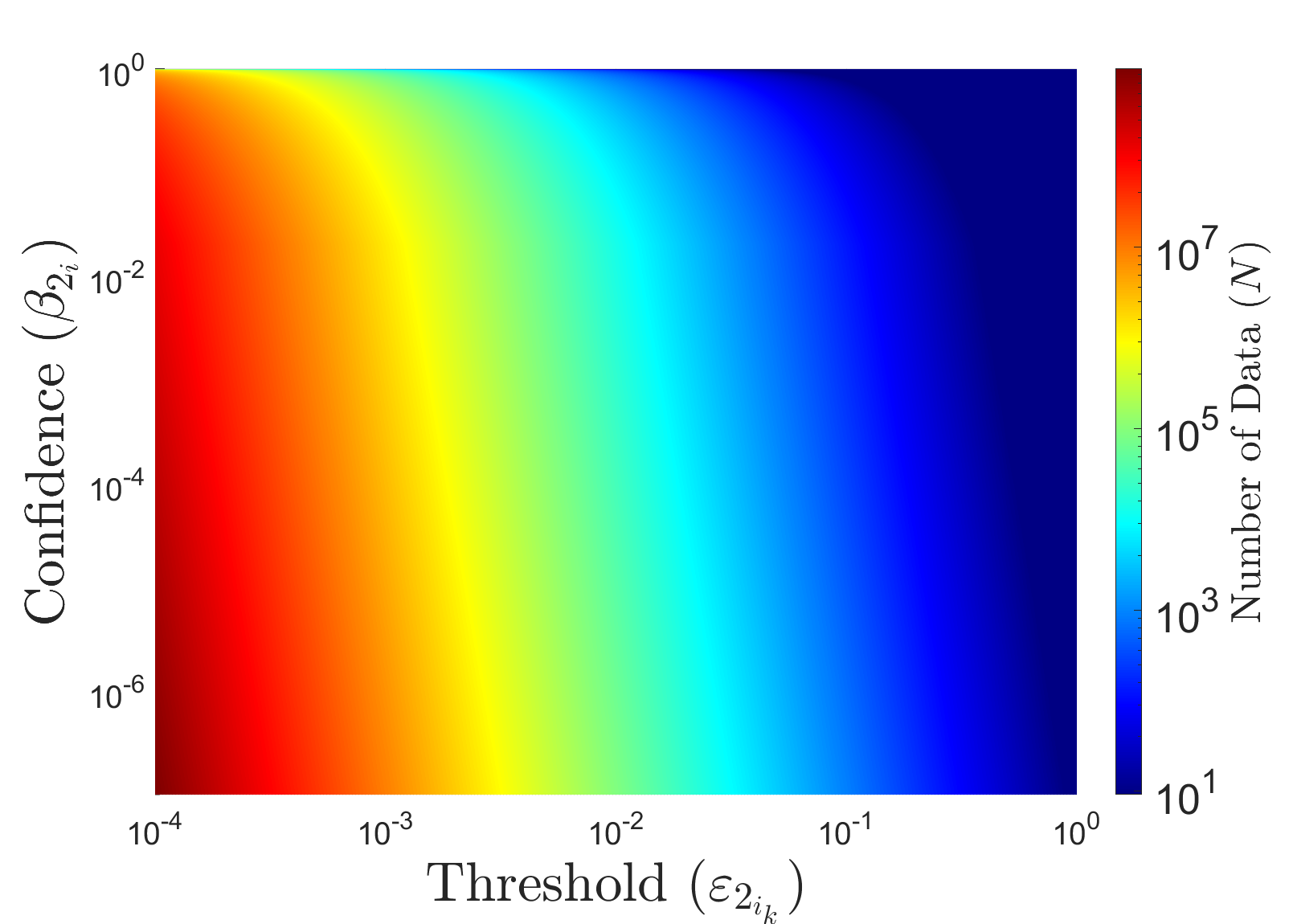}
	\caption{Required number of data, represented by `colour bar', in terms of the threshold $\varepsilon_{2_{i_k}}$ and the confidence $\beta_{2_i}$. Plot is in the logarithmic scale. The required number of data decreases by increasing either the threshold $\varepsilon_{2_{i_k}}$ or the confidence $\beta_{2_i}$.}
	\label{Fig8}
\end{figure}

{\bf Case Study: Vehicle Platooning (computational complexity analysis).} In order to provide a more practical analysis on the computational complexity based on number of collected data required for solving the SOP$_\varsigma$ in~\eqref{SOP_2}, we plot in Fig.~\ref{Fig8} the required number of data in terms of the threshold $\varepsilon_{2_{i_k}}$ and the confidence parameter $\beta_{2_i}$ according to~\eqref{Min_Data} for each vehicle. As it can be observed, the required number of data decreases by increasing either the threshold $\varepsilon_{2_{i_k}}$ or $\beta_{2_i}$.

\section{Discussion}
In this work, we proposed a compositional data-driven scheme for the formal estimation of collision risks for stochastic multi-agent AVs while providing an a-priori guaranteed confidence on our estimation. We first reformulated the original collision risk problem as a robust optimization program (ROP), and then provided a scenario optimization program (SOP) corresponding to the original ROP by collecting finite numbers of data from trajectories of each agent. We then built a probabilistic relation between the optimal value of SOP and that of ROP, and accordingly, constructed a sub-barrier certificate for each unknown agent based on the number of data and a required level of confidence. By leveraging a compositional technique based on small-gain reasoning, we quantified the collision risk for the multi-agent AVs with some level of confidence based on constructed sub-barrier certificates of individual agents. In the case that the compositionality condition is not satisfied, we proposed a relaxed-version of compositional results without requiring any compositionality conditions. We finally presented our techniques for \emph{non-stochastic} multi-agent AVs. We demonstrated the effectiveness of our proposed results by applying them to a vehicle platooning in a network of $100$ vehicles.

\bibliographystyle{IEEEtran}
\bibliography{biblio}

\begin{thebibliography}{10}
\providecommand{\url}[1]{#1}
\csname url@samestyle\endcsname
\providecommand{\newblock}{\relax}
\providecommand{\bibinfo}[2]{#2}
\providecommand{\BIBentrySTDinterwordspacing}{\spaceskip=0pt\relax}
\providecommand{\BIBentryALTinterwordstretchfactor}{4}
\providecommand{\BIBentryALTinterwordspacing}{\spaceskip=\fontdimen2\font plus
\BIBentryALTinterwordstretchfactor\fontdimen3\font minus
  \fontdimen4\font\relax}
\providecommand{\BIBforeignlanguage}[2]{{%
\expandafter\ifx\csname l@#1\endcsname\relax
\typeout{** WARNING: IEEEtran.bst: No hyphenation pattern has been}%
\typeout{** loaded for the language `#1'. Using the pattern for}%
\typeout{** the default language instead.}%
\else
\language=\csname l@#1\endcsname
\fi
#2}}
\providecommand{\BIBdecl}{\relax}
\BIBdecl

\bibitem{Hou2013model}
Z.~Hou and Z.~Wang, ``From model-based control to data-driven control: Survey,
  classification and perspective,'' \emph{Information Sciences}, vol. 235, pp.
  3--35, 2013.

\bibitem{chouhan2020formal}
A.~P. Chouhan and G.~Banda, ``Formal verification of heuristic autonomous
  intersection management using statistical model checking,'' \emph{Sensors},
  vol.~20, no.~16, p. 4506, 2020.

\bibitem{paigwar2020probabilistic}
A.~Paigwar, E.~Baranov, A.~Renzaglia, C.~Laugier, and A.~Legay, ``Probabilistic
  collision risk estimation for autonomous driving: Validation via statistical
  model checking,'' in \emph{2020 IEEE Intelligent Vehicles Symposium (IV)},
  2020, pp. 737--743.

\bibitem{barbier2019validation}
M.~Barbier, A.~Renzaglia, J.~Quilbeuf, L.~Rummelhard, A.~Paigwar, C.~Laugier,
  A.~Legay, J.~Iba{\~n}ez-Guzm{\'a}n, and O.~Simonin, ``Validation of
  perception and decision-making systems for autonomous driving via statistical
  model checking,'' in \emph{2019 IEEE Intelligent Vehicles Symposium (IV)},
  2019, pp. 252--259.

\bibitem{an2020uncertainty}
D.~An, J.~Liu, M.~Zhang, X.~Chen, M.~Chen, and H.~Sun, ``Uncertainty modeling
  and runtime verification for autonomous vehicles driving control: A machine
  learning-based approach,'' \emph{Journal of Systems and Software}, vol. 167,
  2020.

\bibitem{barbot2017statistical}
B.~Barbot, B.~B{\'e}rard, Y.~Duplouy, and S.~Haddad, ``Statistical
  model-checking for autonomous vehicle safety validation,'' in
  \emph{Conference SIA Simulation Num{\'e}rique}, 2017.

\bibitem{prajna2004safety}
S.~Prajna and A.~Jadbabaie, ``Safety verification of hybrid systems using
  barrier certificates,'' in \emph{Proceedings of the International Conference
  on Hybrid Systems: Computation and Control (HSCC)}, 2004, pp. 477--492.

\bibitem{prajna2007framework}
S.~Prajna, A.~Jadbabaie, and G.~J. Pappas, ``A framework for worst-case and
  stochastic safety verification using barrier certificates,'' \emph{TAC},
  vol.~52, no.~8, pp. 1415--1428, 2007.

\bibitem{borrmann2015control}
U.~Borrmann, L.~Wang, A.~D. Ames, and M.~Egerstedt, ``Control barrier
  certificates for safe swarm behavior,'' \emph{IFAC-PapersOnLine}, vol.~48,
  no.~27, pp. 68--73, 2015.

\bibitem{wang2017safety}
L.~Wang, A.~D. Ames, and M.~Egerstedt, ``Safety barrier certificates for
  collisions-free multirobot systems,'' \emph{IEEE Transactions on Robotics},
  vol.~33, no.~3, pp. 661--674, 2017.

\bibitem{ames2019control}
A.~D. Ames, S.~Coogan, M.~Egerstedt, G.~Notomista, K.~Sreenath, and P.~Tabuada,
  ``Control barrier functions: Theory and applications,'' in \emph{Proceedings
  of the 18th European Control Conference (ECC)}, 2019, pp. 3420--3431.

\bibitem{lopez2020robust}
B.~T. Lopez, J.~E. Slotine, and J.~P. How, ``Robust adaptive control barrier
  functions: An adaptive and data-driven approach to safety,'' \emph{IEEE
  Control Systems Letters}, vol.~5, no.~3, pp. 1031--1036, 2020.

\bibitem{folkestad2020data}
C.~Folkestad, Y.~Chen, A.~D. Ames, and J.~W. Burdick, ``Data-driven
  safety-critical control: synthesizing control barrier functions with koopman
  operators,'' \emph{IEEE Control Systems Letters}, vol.~5, no.~6, pp.
  2012--2017, 2020.

\bibitem{robey2021learning}
A.~Robey, L.~Lindemann, S.~Tu, and N.~Matni, ``Learning robust hybrid control
  barrier functions for uncertain systems,'' \emph{IFAC-PapersOnLine}, vol.~54,
  no.~5, pp. 1--6, 2021.

\bibitem{verginis2021safety}
C.~K. Verginis, F.~Djeumou, and U.~Topcu, ``Safety-constrained learning and
  control using scarce data and reciprocal barriers,'' \emph{arXiv:2105.06526},
  2021.

\bibitem{zhang2010safety}
Z.~Zhang, L.and~She, S.~Ratschan, H.~Hermanns, and E.~M. Hahn, ``Safety
  verification for probabilistic hybrid systems,'' in \emph{CAV}, 2010, pp.
  196--211.

\bibitem{yang2020efficient}
Z.~Yang, M.~Wu, and W.~Lin, ``An efficient framework for barrier certificate
  generation of uncertain nonlinear hybrid systems,'' \emph{NAHS}, vol.~36,
  2020.

\bibitem{M.Ahmadi}
M.~Ahmadi, B.~Wu, H.~Lin, and U.~Topcu, ``Privacy verification in {POMDP}s via
  barrier certificates,'' in \emph{Proceedings of the 57th IEEE Conference on
  Decision and Control (CDC)}, 2018, pp. 5610--5615.

\bibitem{ahmadi2019safe}
M.~Ahmadi, A.~Singletary, J.~W. Burdick, and A.~D. Ames, ``Safe policy
  synthesis in multi-agent {POMDPs} via discrete-time barrier functions,'' in
  \emph{Proceedings of the 58th Conference on Decision and Control (CDC)},
  2019, pp. 4797--4803.

\bibitem{santoyo2019verification}
C.~Santoyo, M.~Dutreix, and S.~Coogan, ``Verification and control for
  finite-time safety of stochastic systems via barrier functions,'' in
  \emph{Proceedings of the IEEE Conference on Control Technology and
  Applications}, 2019, pp. 712--717.

\bibitem{clark2019control}
A.~Clark, ``Control barrier functions for complete and incomplete information
  stochastic systems,'' in \emph{Proceedings of the American Control Conference
  (ACC)}, 2019, pp. 2928--2935.

\bibitem{Pushpak2019}
P.~Jagtap, S.~Soudjani, and M.~Zamani, ``Formal synthesis of stochastic systems
  via control barrier certificates,'' \emph{IEEE Transactions on Automatic
  Control}, vol.~66, no.~7, pp. 3097--3110, 2020.

\bibitem{Lavaei_TAC22}
M.~Anand, A.~Lavaei, and M.~Zamani, ``From small-gain theory to compositional
  construction of barrier certificates for large-scale stochastic systems,''
  \emph{IEEE Transactions on Automatic Control}, 2022.

\bibitem{AmyAutomatica2020_J}
A.~Nejati, S.~{Soudjani}, and M.~Zamani, ``Compositional construction of
  control barrier functions for continuous-time stochastic hybrid systems,''
  \emph{Automatica}, 2022.

\bibitem{Amy_LCSS20}
------, ``Compositional construction of control barrier certificates for
  large-scale stochastic switched systems,'' \emph{IEEE Control Systems
  Letters}, vol.~4, no.~4, pp. 845--850, 2020.

\bibitem{Niloo_TCNS}
N.~Jahanshahi, A.~{Lavaei}, and M.~Zamani, ``Compositional construction of
  safety controllers for networks of continuous-space {POMDPs},'' \emph{IEEE
  Transactions on Control of Network Systems}, 2022.

\bibitem{Lavaei_ACC2022}
A.~Lavaei, S.~Soudjani, and E.~Frazzoli, ``Safety barrier certificates for
  stochastic hybrid systems,'' in \emph{Proceedings of the Annual American
  Control Conference}, 2022.

\bibitem{Lavaei_Survey}
A.~Lavaei, S.~Soudjani, A.~Abate, and M.~Zamani, ``Automated verification and
  synthesis of stochastic hybrid systems: A survey,'' \emph{Automatica}, 2022.

\bibitem{Ali_ADHS21}
A.~Salamati, A.~Lavaei, S.~Soudjani, and M.~Zamani, ``Data-driven safety
  verification of stochastic systems via barrier certificates,''
  \emph{Proceedings of the 7th IFAC Conference on Analysis and Design of Hybrid
  Systems (ADHS)}, vol.~54, no.~5, pp. 7--12, 2021.

\bibitem{sadraddini2017provably}
S.~Sadraddini, S.~Sivaranjani, V.~Gupta, and C.~Belta, ``Provably safe cruise
  control of vehicular platoons,'' \emph{IEEE Control Systems Letters}, vol.~1,
  no.~2, pp. 262--267, 2017.

\bibitem{1967stochastic}
H.~J. Kushner, \emph{Stochastic Stability and Control}, ser. Mathematics in
  Science and Engineering.\hskip 1em plus 0.5em minus 0.4em\relax Elsevier
  Science, 1967.

\bibitem{Lavaei_IFAC20}
M.~Anand, A.~Lavaei, and M.~Zamani, ``Compositional construction of control
  barrier certificates for large-scale interconnected stochastic systems,''
  vol.~53, no.~2, pp. 1862--1867, 2020.

\bibitem{saw1984chebyshev}
J.~G. Saw, M.~C. Yang, and T.~C. Mo, ``Chebyshev inequality with estimated mean
  and variance,'' \emph{The American Statistician}, vol.~38, no.~2, pp.
  130--132, 1984.

\bibitem{esfahani2014performance}
P.~Mohajerin~Esfahani, T.~Sutter, and J.~Lygeros, ``Performance bounds for the
  scenario approach and an extension to a class of non-convex programs,''
  \emph{IEEE Transactions on Automatic Control}, vol.~60, no.~1, pp. 46--58,
  2014.

\bibitem{wood1996estimation}
G.~Wood and B.~Zhang, ``Estimation of the {L}ipschitz constant of a function,''
  \emph{Journal of Global Optimization}, vol.~8, no.~1, pp. 91--103, 1996.

\end{thebibliography}

\newpage
\section{Appendix}

\subsection{Additional Lemmas}

\begin{lemma}\label{Lemma:2}
	Consider a linear agent $\mathcal A_i\!: x_i(k+1)=A_ix_i(k)+ D_iw_i(k)+R_i\varsigma_i(k),$ with $A_i, R_i\in\mathbb{R}^{n_i\times n_i}, D_i\in\mathbb{R}^{p_i\times n_i}$, and $\varsigma_i(\cdot) \sim\mathcal N(0, \mathds{I}_n)$. Let $\Vert A_i\Vert \leq \mathscr L_{A_i}, \Vert D_i\Vert \leq \mathscr L_{D_i}$, where $\mathscr L_{A_i}, \mathscr L_{D_i}\in\mathbb R_{0}^+$. Then $\mathscr{L}_{g_{i_k}}$ for a quadratic SBC in the form of $x_i^\top P_ix_i,$ with a positive-definite matrix $P_i\in\mathbb{R}^{n_i\times n_i}$, is computed as 
	\begin{align*}
	\mathscr{L}_{g_{i_k}} &=\max \Big\{\mathscr{L}_{g_{i_1}},\mathscr{L}_{g_{i_2}},\mathscr{L}_{g_{i_3}}, \mathscr{L}_{g_{i_4}},\mathscr{L}_{g_{i_{5_k}}}\Big\} = \max\big\{2s_i(\lambda_{\max}(P_i)+\mathscr L_{\alpha_i}),(\mathscr{L}_{g_{i_{5_{k x}}}}^2 + \mathscr{L}_{g_{i_{5_{w}}}}^2)^{\frac{1}{2}}\big\},
	\end{align*} 
	with 
	\begin{align*} \mathscr{L}_{g_{i_{5_{k x}}}} &=2\lambda_{\max}(P_i)(s_i(\mathscr L_{A_i}^2+\kappa_i) + s'_i\mathscr L_{A_i}\mathscr L_{D_i}),\\ \mathscr{L}_{g_{i_{5_{w}}}}&=2\lambda_{\max}(P_i)(s'_i\mathscr L_{D_i}^2 + s_i\mathscr L_{A_i}\mathscr L_{D_i}) + 2\mathscr L_{\rho_i}s'_i,
	\end{align*} 
	where $\mathscr L_{\alpha_i},\mathscr L_{\rho_i},s_i,s'_i,$ are, respectively, upper bounds on $\alpha_i,\rho_i$, and the norm of $x_i, w_i$, \emph{i.e.,} $\alpha_i \leq \mathscr L_{\alpha_i}\in\mathbb R_{0}^+, \rho_i \leq \mathscr L_{\rho_i}\in\mathbb R_{0}^+, \Vert x_i\Vert \leq s_i\in\mathbb R_{0}^+, \forall x_i\in X_i$, and $\Vert w_i\Vert \leq s'_i\in\mathbb R_{0}^+, \forall w_i\in W_i$.
\end{lemma}

\begin{remark}\label{Gerschgorin}
	One needs to know upper bounds for $\lambda_{\max}(P_i),\mathscr L_{\alpha_i},\mathscr L_{\rho_i}$ in order to compute $\mathscr{L}_{g_{i_k}}$, and accordingly, the required number of data as Step 3 in Algorithm~\ref{Alg:1}. The pre-assumed upper bounds should be then enforced as some additional conditions during solving the $\text{SOP}_\varsigma$~\eqref{SOP_2} as mentioned in Step 4 of Algorithm~\ref{Alg:1}.
\end{remark}

Similarly, we provide another lemma for the computation of $\mathscr{L}_{g_{i_k}}$ but for \emph{nonlinear} agents $\mathcal A_i$.

\begin{lemma}\label{Lemma:3}
	Consider nonlinear agents $\mathcal A_i\!\!: x_i(k+1)=f_i(x_i(k),w_i(k)) +R_i\varsigma_i(k)$ with $\varsigma_i(\cdot) \sim\mathcal N(0, \mathds{I}_n)$. Let $\Vert f_i(x_i,w_i)\Vert \leq \mathscr L_i, \Vert \partial_{x_i}f_i(x_i,w_i) \Vert = \Vert \frac{\partial f_i(x_i,w_i)}{\partial x_i}\Vert \leq \mathscr L_{x_i}, \Vert \partial_{w_i}f_i(x_i,w_i) \Vert = \Vert \frac{\partial f_i(x_i,w_i)}{\partial w_i}\Vert \leq \mathscr L_{w_i}$, where $\mathscr L_i, \mathscr L_{x_i}, \mathscr L_{w_i} \in \mathbb R_{0}^+$. Then $\mathscr{L}_{g_{i_k}}$ for a quadratic SBC in the form of $x_i^\top P_ix_i$ with a positive-definite matrix $P_i\in\mathbb{R}^{n_i\times n_i}$, is computed as 
	\begin{align*}
	\mathscr{L}_{g_{i_k}} &=\max \big\{\mathscr{L}_{g_{i_1}},\mathscr{L}_{g_{i_2}},\mathscr{L}_{g_{i_3}}, \mathscr{L}_{g_{i_4}},\mathscr{L}_{g_{i_{5_k}}}\big\} = \max\big\{2s_i(\lambda_{\max}(P_i) + \mathscr L_{\alpha_i}),(\mathscr{L}_{g_{i_{5_{k x}}}}^2 + \mathscr{L}_{g_{i_{5_w}}}^2)^{\frac{1}{2}}\big\},
	\end{align*}
	with
	\begin{align*}
	\mathscr{L}_{g_{i_{5_{k x}}}} &=2\lambda_{\max}(P_i)(\mathscr L_{i}\mathscr L_{x_i}+\kappa_is_i),\\
	\mathscr{L}_{g_{i_{5_w}}} &= 2\lambda_{\max}(P_i)\mathscr L_{i}\mathscr L_{w_i} + 2\mathscr L_{\rho_i}s'_i,\end{align*}
	where $\mathscr L_{\alpha_i},\mathscr L_{\rho_i},s_i,s'_i,$ are, respectively, upper bounds on $\alpha_i,\rho_i,$ and the norm of $x_i, w_i$, \emph{i.e.,} $\alpha_i \leq \mathscr L_{\alpha_i}\in\mathbb R_{0}^+, \rho_i \leq \mathscr L_{\rho_i}\in\mathbb R_{0}^+, \Vert x_i\Vert \leq s_i\in\mathbb R_{0}^+, \forall x_i\in X_i$, and $\Vert w_i\Vert \leq s'_i\in\mathbb R_{0}^+, \forall w_i\in W_i$.
\end{lemma}

\subsection{Proofs of Statements}

\begin{proof}\textbf{(Theorem~\ref{Thm:2})}
	Based on~\cite[Theorem 4.1, 4.3]{esfahani2014performance}, the
	probabilistic distance between optimal values of
	ROP and $\text{SOP}_\text{N}$ can be formally lower bounded by\footnote{One can readily verify that $\eta^*_{R_i}$ is always bigger than or equal to $\eta^*_{N_i}$ because $\eta^*_{R_i}$ is computed for infinitely-many constraints, whereas $\eta^*_{N_i}$ is computed only for finitely many of them.}
	\begin{align}\notag
	\PP^{N_i}\Big\{0\leq\eta^*_{R_i} - \eta^*_{N_i}\leq\varepsilon_{1_i}\Big\}\geq 1-\beta_{2_i},
	\end{align}
	provided that $$N_i\geq N_i\big(g_{i_k}(\frac{\varepsilon_{1_i}}{\mathrm L_{\mathrm {SP}_i}\mathscr{L}_{g_{i_k}}}),\beta_{2_i}\big),$$ where
	$g_{i_k}: [0,1]\rightarrow [0,1]$ is given by
	\begin{align}\notag
	g_{i_k}(s) = s^{n_i + p_i}, \quad \forall s \in  [0,1],
	\end{align}
	and $\mathrm{L}_{\mathrm{SP}_i}$ is a Slater constant as defined in~\cite[equation (5)]{esfahani2014performance}. Since the original ROP in~\eqref{ROP} can be cast as a $\min$-$\max$ optimization problem, the Slater constant  $\mathrm{L}_{\mathrm{SP}_i}$ can be selected as 1~\cite[Remark 3.5]{esfahani2014performance}.
	
	\noindent
	One can readily conclude that $\eta^*_{N_i}\leq\eta^*_{R_i}\leq\eta^*_{N_i} + \varepsilon_{1_i}$ with a confidence of at least $1-\beta_{2_i}$. From Lemma~\ref{Lemma:1}, we have $\eta^*_{N_i} \leq \eta^*_{\varsigma _i}$ with a confidence of at least  $1-\beta_{1_i}$. Let us now define events $\mathcal E_1 :=\{ \eta^*_{N_i} \leq \eta^*_{\varsigma _i}\}$ and $\mathcal E_2 :=\{\eta^*_{R_i}\leq\eta^*_{N_i} + \varepsilon_{1_i}\}$, where $\PP\big\{\mathcal E_1\big\}\geq1-\beta_{1_i}$ and $\PP\big\{\mathcal E_2\big\}\geq1-\beta_{2_i}$. From the above derivations, one has $\eta^*_{R_i}\leq\eta^*_{N_i} + \varepsilon_{1_i} \leq \eta^*_{\varsigma_i} + \varepsilon_{1_i}$. Since $\eta^*_{\varsigma_i} + \varepsilon_{1_i} \leq 0$, it implies that $\eta^*_{R_i} \leq 0$. We are now computing the concurrent occurrence of  events $\mathcal E_1$ and $\mathcal E_2$, namely $\PP\big\{\mathcal E_1\cap \mathcal E_2\big\}$:
	\begin{align}\label{Eq:16}
	&\PP\big\{\mathcal E_1\cap \mathcal E_2\big\}=1-\PP\big\{\bar {\mathcal E_1}\cup \bar {\mathcal E_2}\big\},
	\end{align}
	where $\bar {\mathcal E_1}$ and $\bar {\mathcal E_2}$ are the complement of $\mathcal E_1$ and $\mathcal E_2$, respectively. Since
	\begin{align*}
	&\PP\big\{\bar {\mathcal E_1}\cup \bar {\mathcal E_2}\big\}\leq\PP\big\{\bar {\mathcal E_1}\big\}+\PP\big\{\bar {\mathcal E_2}\big\},
	\end{align*}
	and by leveraging \eqref{Eq:16}, one can readily conclude that 
	\begin{align}
	\PP\big\{\mathcal E_1\cap \mathcal E_2\big\}\geq 1-\PP\big\{\bar {\mathcal E_1}\big\}-\PP\big\{\bar {\mathcal E_2}\big\}\nonumber
	\geq 1-\beta_{1_i}-\beta_{2_i}.
	\end{align}
	Therefore, the solution $\Theta^*_i$ via solving $\text{SOP}_\varsigma$ in~\eqref{SOP_2} is a feasible solution for ROP~\eqref{ROP} with a confidence of at least $1-\beta_{1_i}-\beta_{2_i}$, which completes the proof. 
\end{proof}

\begin{proof}\textbf{(Lemma~\ref{Lemma:2})}
	We first compute Lipschitz constants of $\mathscr{L}_{g_{i_1}}$-$\mathscr{L}_{g_{i_4}}$ with respect to $x$ and $\mathscr{L}_{g_{i_{5_k}}}$ with respect to $x_i$ and $w_i$, and then take the maximum among them. By defining 
	\begin{align}\notag
	\mathscr{L}_{g_{i_{5_{k x}}}}\!\!:\left\{
	\hspace{-0.5mm}\begin{array}{l}\max\limits_{x_i\in X_i} \quad\quad\!\!\Vert\frac{\partial g_{i_{5_k}}}{\partial x_i}\Vert, \\
	\,\,\! \text{s.t.} \quad \quad \,\,\,\,\Vert x_i\Vert \leq s_i,  \Vert w_i\Vert \leq s_i',\end{array}\right.
	\end{align}
	one has 
	\begin{align*}
	\mathscr{L}_{g_{i_{5_{k x}}}}&= \max\limits_{x_i\in X_i}\Vert 2(A_i^\top P_iA_i - \kappa_{i_k}P_i)x_i + 2A_i^\top P_iD_iw_i\Vert\\
	&\leq\max\limits_{x_i\in X_i}\Vert 2(A_i^\top P_iA_i - \kappa_{i_k}P_i)\Vert \Vert x_i\Vert + \Vert 2A_i^\top P_iD_i\Vert \Vert w_i\Vert\\
	&\leq 2s_i(\Vert A_i^\top P_iA_i\Vert + \kappa_{i_k}\Vert P_i\Vert) + 2s'_i\Vert A_i\Vert\Vert P_i\Vert\Vert D_i\Vert\\
	&\leq 2s_i(\Vert P_i\Vert \Vert A_i\Vert^2+ \kappa_{i_k}\Vert P_i\Vert) + 2s'_i\Vert A_i\Vert\Vert P_i\Vert\Vert D_i\Vert\\
	&\leq 2\lambda_{\max}(P_i)(s_i(\mathscr L_{A_i}^2+\kappa_{i_k}) + s'_i\mathscr L_{A_i}\mathscr L_{D_i}).
	\end{align*}
	Similarly, be defining \begin{align}\notag
	\mathscr{L}_{g_{i_{5_{w}}}}\!\!:\left\{
	\hspace{-0.5mm}\begin{array}{l}\max\limits_{w_i\in W_i} \quad\quad\!\!\!\Vert\frac{\partial g_{i_{5}}}{\partial w_i}\Vert, \\
	\,\,\! \text{s.t.} \quad \quad \,\,\,\,\,\Vert x_i\Vert \leq s_i,  \Vert w_i\Vert \leq s_i',\end{array}\right.
	\end{align}
	one has 
	\begin{align*}
	\mathscr{L}_{g_{i_{5_w}}}&= \max\limits_{w_i\in W_i}\Vert 2D_i^\top P_iD_iw_i + 2D_i^\top P_iA_ix_i - 2\rho_iw_i\Vert\\
	&\leq\max\limits_{w_i\in W_i}2\Vert D_i^\top P_iD_i\Vert \Vert w_i\Vert + 2\Vert D_i^\top P_iA_i\Vert \Vert x_i\Vert + 2\rho_i\Vert w_i\Vert \\
	&\leq 2s'_i\Vert D_i^\top P_iD_i\Vert + 2s_i\Vert D_i\Vert\Vert P_i\Vert\Vert A_i\Vert+ 2\mathscr L_{\rho_i}s'_i\\
	&\leq 2\lambda_{\max}(P_i)(s'_i\mathscr L_{D_i}^2 + s_i\mathscr L_{A_i}\mathscr L_{D_i}) + 2\mathscr L_{\rho_i}s'_i.
	\end{align*}
	
	Then $\mathscr{L}_{g_{i_{5_k}}}= (\mathscr{L}_{g_{i_{5_{k x}}}}^2 + \mathscr{L}_{g_{i_{5_w}}}^2)^{\frac{1}{2}}$. Similarly for $g_{i_1}$-$g_{i_4}$, we have $\mathscr{L}_{g_{i_1}} \!=\! \mathscr{L}_{g_{i_3}} \!=\! \mathscr{L}_{g_{i_4}} \!\leq\! 2s_i\lambda_{\max}(P_i),~ \mathscr{L}_{g_{i_2}} \!\leq\! 2s_i(\lambda_{\max}(P_i) + \mathscr L_{\alpha_i})$. Then $\mathscr{L}_{g_i} \!=\!\max \big\{\mathscr{L}_{g_{i_1}},\mathscr{L}_{g_{i_2}},\mathscr{L}_{g_{i_3}}, \mathscr{L}_{g_{i_4}},\mathscr{L}_{g_{i_{5_k}}}\big\} = \max\big\{2s_i(\lambda_{\max}(P_i)+\mathscr L_{\alpha_i}),(\mathscr{L}_{g_{i_{5_{k x}}}}^2 + \mathscr{L}_{g_{i_{5_w}}}^2)^{\frac{1}{2}}\big\}$, which completes the proof.
\end{proof}

\begin{proof}\textbf{(Lemma~\ref{Lemma:3})}
	By defining 
	\begin{align*}
	\mathscr{L}_{g_{i_{5_{k x}}}}\!:&\left\{
	\hspace{-0.5mm}\begin{array}{l}\max\limits_{x_i\in X_i} \quad\quad\!\!\Vert\frac{\partial g_{i_{5_k}}}{\partial x_i}\Vert, \\
	\,\,\! \text{s.t.} \quad \quad \,\,\,\,\Vert x_i\Vert \leq s_i,  \Vert w_i\Vert \leq s_i',\end{array}\right.\\
	\mathscr{L}_{g_{i_{5_w}}}\!:&\left\{
	\hspace{-0.5mm}\begin{array}{l}\max\limits_{w_i\in W_i} \quad\quad\!\!\Vert\frac{\partial g_{i_{5}}}{\partial w_i}\Vert, \\
	\,\,\! \text{s.t.} \quad \quad \,\,\,\,\,\Vert x_i\Vert \leq s_i,  \Vert w_i\Vert \leq s_i',\end{array}\right.
	\end{align*}
	one has $\mathscr{L}_{g_{i_{{5_{k x}}}}}\leq 2\lambda_{\max}(P_i)(\mathscr L_{i}\mathscr L_{x_i}+\kappa_{i_k}s_i), \mathscr{L}_{g_{i_{5_w}}}\leq  2\lambda_{\max}(P_i)\mathscr L_{i}\mathscr L_{w_i} + 2\mathscr L_{\rho_i}s'_i$. Then $\mathscr{L}_{g_{i_{5_{k}}}}= (\mathscr{L}_{g_{i_{5_{k x}}}}^2 + \mathscr{L}_{g_{i_{5_w}}}^2)^{\frac{1}{2}}$. Similarly for $g_{i_1}$-$g_{i_4}$, we have $\mathscr{L}_{g_{i_1}} = \mathscr{L}_{g_{i_3}} = \mathscr{L}_{g_{i_4}} \leq 2s_i\lambda_{\max}(P_i),~ \mathscr{L}_{g_{i_2}} \leq 2s_i(\lambda_{\max}(P_i) + \mathscr L_{\alpha_i})$. Then $\mathscr{L}_{g_{i_k}} =\max \big\{\mathscr{L}_{g_{i_1}},\mathscr{L}_{g_{i_2}},\mathscr{L}_{g_{i_3}}, \mathscr{L}_{g_{i_4}},\mathscr{L}_{g_{i_{5_k}}}\big\} \!=\! \max\big\{2s_i(\lambda_{\max}(P_i) + \mathscr L_{\alpha_i}),(\mathscr{L}_{g_{i_{5_{k x}}}}^2 + \mathscr{L}_{g_{i_{5_w}}}^2)^{\frac{1}{2}}\big\}$, which completes the proof.
\end{proof}

\begin{proof}\textbf{(Theorem~\ref{Thm:3})}
	We first show that conditions~\eqref{Eq:6_1}-\eqref{Eq:6_2} hold. For any $x\Let[x_{1};\ldots;x_{M}] \in X_0 = \prod_{i=1}^{M} X_{0_i} $, we have
	\begin{align}\notag
	\mathcal B(q,x) = \sum_{i=1}^{M}\mathcal B_i(q_i,x_i)\leq \sum_{i=1}^{M}\gamma_i = \gamma,
	\end{align} 
	and similarly for any $x\Let[x_{1};\ldots;x_{N}] \in X_c = \prod_{i=1}^{M} X_{c_i} $, one has
	\begin{align}\notag
	\mathcal B(q,x) = \sum_{i=1}^{M}\mathcal B_i(q_i,x_i)\geq \sum_{i=1}^{M}\lambda_i = \lambda,
	\end{align} 
	satisfying conditions~\eqref{Eq:6_1}-\eqref{Eq:6_2} with $\gamma = \sum_{i=1}^{M}\gamma_i$ and $\lambda = \sum_{i=1}^{M}\lambda_i$. Note that $\lambda > \gamma$ according to~\eqref{Eq:43}.
	
	Now we show that condition~\eqref{Eq:6_3} holds, as well. By employing condition~\eqref{Eq:8_1} and compositionality condition $\mathds{1}_M^\top(-\Lambda+\Delta)< 0$, one can obtain the chain of inequalities in \eqref{Eq:17}. By defining
	\begin{align}\notag
	\kappa s &\Let \max\Big\{s + \mathds{1}_M^\top(-\Lambda+\Delta)\bar {\mathcal B}(q,x)\,\big|\, \mathds{1}_M^\top\bar {\mathcal B}(q,x)=s\Big\}, \\\label{Eq:18}
	\psi&\Let\sum_{i=1}^M\psi_i,
	\end{align}
	where $\bar {\mathcal B}(q,x)=\big[\mathcal B_1(q_1,x_1);\ldots;\mathcal B_M(q_M,x_M)\big]$, condition \eqref{Eq:6_3} is also satisfied. We now show that $ \kappa = 1 + \pi$ and $0<\kappa<1$. Since $\mathds{1}_M^\top(-\Lambda+\Delta) \Let \big[\pi_1;\dots;\pi_M\big]^\top < 0$, and  $\max_{1\leq i\leq M} \pi_i < \pi < 0$ with $\pi \in (-1,0)$, one has
	\begin{align*}
	\kappa s &= s + \mathds{1}_M^\top(-\Lambda+\Delta)\bar {\mathcal B}(q,x) = s + \big[\pi_1;\dots;\pi_M\big]^\top\big[\mathcal B_1(q_1,x_1);\ldots;\mathcal B_M(q_M,x_M)\big]\\
	&= s + \pi_1\mathcal B_1(q_1,x_1) + \dots + \pi_M\mathcal B_M(q_M,x_M)\leq s + \pi\big(\mathcal B_1(q_1,x_1) + \dots + \mathcal B_M(q_M,x_M)\big) = s + \pi s = (1 + \pi) s.
	\end{align*}
	Then $\kappa s \leq (1 + \pi) s$, and accordingly, $\kappa \leq1 + \pi$. Since $\max_{1\leq i\leq M} \pi_i < \pi < 0$ with $\pi \in (-1,0)$, then $0 <\kappa = 1 + \pi <1$, which completes the proof. Note that since each agent $\mathcal A_i$ admits its SBC $\mathcal B_i$ with a confidence of at least $1-\beta_{1_i}-\beta_{2_i}$, one can readily conclude that $\mathcal B$ is a BC for the multi-agent AV $\mathcal A$ with a confidence of at least $1-\sum_{i=1}^{M}\beta_{1_i}-\sum_{i=1}^{M}\beta_{2_i}$.
\end{proof}

\begin{figure*}
	\rule{\textwidth}{0.1pt}
	\begin{align}\notag
	\EE&\Big[\mathcal B(q,f(x,\varsigma))\,\big|\,x \Big]=\EE\Big[\sum_{i=1}^{M}\mathcal B_{i}(q_i,f_i(x_i,w_i,\varsigma_i))\,\big|\,x,w\Big]=\sum_{i=1}^{M}\EE\Big[\mathcal B_{i}(q_i,f_i(x_i,w_i,\varsigma_i))\,\big|\,x,w\Big]\\\notag
	&=\sum_{i=1}^{M}\EE\Big[\mathcal B_{i}(q_i,f_i(x_i,w_i,\varsigma_i))\,\big|\,x_i,w_i\Big]\leq\sum_{i=1}^{M}\big(\kappa_{i}\mathcal B_{i}(q_i,x_i)+\rho_{i}\Vert w_i\Vert^2+ \psi_{i}\big)\\\notag
	&\leq\sum_{i=1}^{M}\big(\kappa_{i}\mathcal B_{i}(q_i,x_i)+\rho_{i}\sum_{j=1,i\neq{j}}^M\Vert w_{ij}\Vert^2+ \psi_{i}\big)=\sum_{i=1}^{M}\big(\kappa_{i}\mathcal B_{i}(q_i,x_i)+\rho_{i}\!\!\!\sum_{j=1,i\neq{j}}^M\Vert x_{j}\Vert^2+ \psi_{i}\big)\\\notag
	&=\sum_{i=1}^{M}\big(\kappa_{i}\mathcal B_{i}(q_i,x_i)+\sum_{j=1,i\neq{j}}^M\rho_{i}\Vert x_{j}\Vert^2+\psi_{i}\big)\leq\sum_{i=1}^{M}\big(\kappa_{i}\mathcal B_{i}(q_i,x_i)+\sum_{j=1,i\neq{j}}^M\frac{\rho_{i}}{\alpha_{j}}\mathcal B_{j}(q_j,x_j)+ \psi_{i}\big)\\\notag
	&=\sum_{i=1}^{M}\mathcal B_i(q_i,x_i) - \sum_{i=1}^M\big(\hat\lambda_i\mathcal B_i(q_i, x_i)+\sum_{j=1,i\neq{j}}^M\hat\delta_{ij}\mathcal B_j(q_j,x_j)+\psi_i\big)\\\label{Eq:17}
	&=\mathcal B(q,x) + \mathds{1}_M^\top(-\Lambda+\Delta)\big[\mathcal B_1(q_1,x_1);\ldots;\mathcal B_M(q_M,x_M)\big]+\sum_{i=1}^M\psi_i\leq (1 + \pi) \mathcal B(q,x) + \psi.
	\end{align}
	\rule{\textwidth}{0.1pt}
\end{figure*}

\begin{proof}\textbf{(Theorem~\ref{Thm:4})}
	According to condition~\eqref{Eq:8_3}, one has $X_{c_i}\subseteq \{x_i\in X_i \,\,\big|\,\, \mathcal B_i(x_i) \ge \lambda_i \}$. Then we have
	\begin{align}\notag
	\PP&\Big\{x_{x_{0_i}}(k)\in X_{c_i} \text{ for } 0\leq k < \mathcal T_i\,\,\big|\,\, x_{0_i} = x_i(0)\Big\}\\\label{Eq:15}
	&\leq\PP\Big\{ \sup_{0\leq k < \mathcal T} \mathcal B_i(x_i(k))\geq \lambda_i \,\,\big|\,\, x_{0_i} = x_i(0) \Big\}.
	\end{align}
	Since
	\begin{align}\label{Eq:8_5}
	\EE&\Big[\mathcal{B}_i(f_i(x_i,w_i,\varsigma_i))\,\big |\, x_i,w_i\Big]\leq\kappa_i\mathcal{B}_i(x_i)+\hat\psi_i, ~\text{with}~  \hat\psi_i= \rho_i \Vert w_i\Vert^2_\infty + \psi_i,
	\end{align}
	the collision risk estimation for each agent $\mathcal A_i$ in~\eqref{Eq:12} follows directly by applying~\cite[Corrolary 2-1, Chapter III]{1967stochastic} to~\eqref{Eq:15} and employing, respectively, conditions~\eqref{Eq:8_5} and~\eqref{Eq:8_2}.
\end{proof}

\begin{proof}\textbf{(Theorem~\ref{Thm:5})}
	Since 
	\begin{align*}
	\PP&\big\{\mathcal A_1\models \varphi_1 \cup\dots\cup\mathcal A_M\models\varphi_M\big\}\leq \PP\big\{\mathcal A_1\models \varphi_1 \big\} + \dots + \PP\big\{\mathcal A_M\models \varphi_M \big\},
	\end{align*}
	we have 
	$$\PP\big\{\mathcal A_1\models \varphi_1 \cup\dots\cup\mathcal A_M\models\varphi_M\big\} \leq \delta_1 + \dots + \delta_M = \sum_{i =1}^M\delta_i.$$ By defining collision events $\mathcal E_i$ and $\mathcal E$ as 
	\begin{align*}
	&\mathcal E_i = \Big\{\PP\big\{\mathcal A_i\models \varphi_i \big\}\leq\delta_i\Big\}, ~~\forall i \in\{1,\dots,M\},\\
	&\mathcal E = \Big\{\PP\big\{\mathcal A_1\models \varphi_1 \cup\dots\cup\mathcal A_M\models\varphi_M \big\}\leq\sum_{i =1}^M\delta_i \Big\},
	\end{align*} 
	one has $\PP\big\{\mathcal E_i\big\}\geq 1-\beta_i, \forall i \in\{1,\dots,M\}$. We now aim at computing a lower bound for $\PP\big\{\mathcal E\big\}$. We have 
	\begin{align}\label{Eq:19}
	\PP&\big\{\bar {\mathcal E}_1\cup\dots\cup \bar {\mathcal E}_M \big\}\leq\PP\big\{\bar {\mathcal E}_1\big\}+\dots+\PP\big\{\bar {\mathcal E}_M\big\}\leq \beta_1 +\dots + \beta_M = \sum_{i =1}^M\beta_i,
	\end{align}
	with $\bar {\mathcal E}_i$ being complements of $\mathcal E_i, \forall i \in\{1,\dots,M\}$. Since $\bar {\mathcal{E}}\subseteq \bar {\mathcal E}_1\cup\dots\cup \bar {\mathcal E}_M$, one has $\PP\big\{\bar {\mathcal{E}}\big\} \leq \PP\big\{\bar {\mathcal E}_1\cup\dots\cup \bar {\mathcal E}_M\big\} $. By employing~\eqref{Eq:19}, one has $\PP\big\{\bar {\mathcal{E}}\big\} \leq \sum_{i =1}^M\beta_i$. Subsequently, $\PP\big\{\mathcal{E}\big\} \geq 1- \sum_{i =1}^M\beta_i$,	
	which completes the proof.
\end{proof}

\begin{proof}\textbf{(Theorem~\ref{Thm:9})}
	We show the statement based on contradiction. Let $x_{x_{0}}$ of $\mathcal A$ start at some $x_{0}\in X_{0}$. Suppose $x_{x_{0}}$ reaches inside $X_{c}$. Based on conditions~\eqref{Eq:6_1}-\eqref{Eq:6_2}, one has $\mathcal B(x(0)) \leq \gamma$ and $\mathcal B(x(k)) \geq \lambda$ for some $k\in \mathbb N$. Since  $\mathcal B(x(\cdot))$ is a BC and by employing
	condition~\eqref{Eq:6_4}, one can conclude that $\lambda \leq \mathcal B(x(k)) \leq \mathcal B(x(0)) \leq \gamma$. This contradicts condition $\lambda > \gamma$, which completes the proof. 
\end{proof}

\begin{proof}\textbf{(Theorem~\ref{Thm:6})}
	According to~\eqref{Eq:14}, since $\mathcal{B}_i(x_i(k+1))-\mathcal{B}_i(x_i(k))\leq \rho_i\Vert w_i\Vert^2_\infty,$
	one can recursively infer that $\mathcal{B}_i(x_i(k))-\mathcal{B}_i(x_i(0))\leq \rho_i\Vert w_i\Vert^2_\infty k.$
	By employing condition~\eqref{Eq:8_2}, we have $\mathcal{B}_i(x_i(k))\leq \gamma_i + \rho_i\Vert w_i\Vert^2_\infty k.$
	Now since $\gamma_i + \rho_i\Vert w_i\Vert^2_\infty\mathcal T_i < \lambda_i$, 
	one can readily conclude that $ \mathcal{B}_i(x_i(k)) < \lambda_i$. From condition~\eqref{Eq:8_3}, one gets $x_i(k) \notin X_{c_i}$ for any $k\in[0,\mathcal T_i)$. This implies that the collision risk is zero for each agent $\mathcal A_i$ for the finite time horizon $\mathcal T_i = \frac{\lambda_i - \gamma_i}{\rho_i\Vert w_i\Vert^2_\infty}$, which completes the proof.
\end{proof}

\end{document}